\documentclass[sigconf]{acmart}

\usepackage{amsmath,amsfonts}
\usepackage{algorithmic}
\usepackage{graphicx}
\usepackage{textcomp}
\usepackage{xcolor}
\usepackage{tcolorbox}
\usepackage{graphicx}
\usepackage{listings}
\usepackage{makecell}
\usepackage{adjustbox}
\usepackage{multirow}
\usepackage{graphicx}
\usepackage{svg}
\usepackage{xspace}
\usepackage{amsmath}
\usepackage{subfigure}
\usepackage{caption}
\usepackage{amsmath}
\usepackage{booktabs}
\usepackage{color, colortbl}
\usepackage{url}
\usepackage{caption} 

\DeclareUnicodeCharacter{2212}{-}
\definecolor{Gray}{gray}{0.9}
\definecolor{green}{RGB}{102,252,102}
\definecolor{ored}{RGB}{255,99,71}
\definecolor{orange}{RGB}{255,165,0}
\definecolor{lightgray}{RGB}{211,211,211}

\usepackage[ruled,vlined,linesnumbered]{algorithm2e}

\SetCommentSty{mycommfont}

\usepackage{tikz}
\usetikzlibrary{shapes}

\newcommand{\tool}{{\texttt{GIFdroid}}\xspace}

\newcommand{\sidong}[1]{\textcolor{blue}{\textbf{Sidong}: #1}}
\newcommand{\chen}[1]{\textcolor{red}{\textbf{Chen}: #1}}

\AtBeginDocument{%
  \providecommand\BibTeX{{%
    \normalfont B\kern-0.5em{\scshape i\kern-0.25em b}\kern-0.8em\TeX}}}

\copyrightyear{2022}
\acmYear{2022}
\setcopyright{acmcopyright}\acmConference[ICSE '22]{44th International Conference on Software Engineering}{May 21--29, 2022}{Pittsburgh, PA, USA}
\acmBooktitle{44th International Conference on Software Engineering (ICSE '22), May 21--29, 2022, Pittsburgh, PA, USA}
\acmPrice{15.00}
\acmDOI{10.1145/3510003.3510048}
\acmISBN{978-1-4503-9221-1/22/05}

\begin{document}

\title{GIFdroid: Automated Replay of Visual Bug Reports for Android Apps}

\author{Sidong Feng}
\affiliation{%
  \institution{Monash University}
  \city{Melbourne}
  \country{Australia}}
\email{sidong.feng@monash.edu}

\author{Chunyang Chen}
\affiliation{%
  \institution{Monash University}
  \city{Melbourne}
  \country{Australia}}
\email{chunyang.chen@monash.edu}
\authornote{Corresponding authors}

\renewcommand{\shortauthors}{Feng et al.}

\begin{abstract}
Bug reports are vital for software maintenance that allow users to inform developers of the problems encountered while using software.
However, it is difficult for non-technical users to write clear descriptions about the bug occurrence.
Therefore, more and more users begin to record the screen for reporting bugs as it is easy to be created and contains detailed procedures triggering the bug.
But it is still tedious and time-consuming for developers to reproduce the bug due to the length and unclear actions within the recording.
To overcome these issues, we propose \tool, a light-weight approach to automatically replay the execution trace from visual bug reports.
\tool adopts image processing techniques to extract the keyframes from the recording, map them to states in GUI Transitions Graph, and generate the execution trace of those states to trigger the bug.
Our automated experiments and user study demonstrate its accuracy, efficiency, and usefulness of the approach.
%To lower the barrier for documenting the bug, video recordings are becoming a more popular artifact.
%However, replaying the video based bug report presents serious challenges, due to the uncertainty of steps to the entry frame.
%To address these challenges, this paper introduces \tool, a novel approach for automatically replaying the execution trace from visual bug reports.
%\tool uses image processing techniques to extract the keyframes of bug reports, map GUI to states in GUI Transitions Graph (UTG), and generate the execution trace to automatically repeat the bug trigger.
%Our comprehensive experiments unveil that our approach can accurately replay the execution traces for visual bug recordings.
\end{abstract}





\maketitle

\section{Introduction}
% Previous studies have also shown that the information most useful to developers is often the most diï¬cult for reporters to provide and thatt he lack of this information is a major reason behind non-reproducible bug reports

% Typically, developers are provided with a bug report that
% contains data about the failure, such as memory dumps and,
% in the best case, some additional information provided by the
% user. However, this data is usually insufficient for recreating
% the problem, as recently reported in a survey conducted among
% developers of the Apache, Eclipse, and Mozilla projects. Even
% more advanced approaches for gathering field data and help
% in-house debugging tend to collect either too little information,
% which results in inexpensive but often ineffective techniques, or
% too much information, which makes the techniques effective but
% too costly.
% Our results are promising and lead to several
% findings, some of which unexpected.

% \chen{Need a paragraph to tell the importance of bug reports and also how to write a good bug reports.}
% \chen{2) Please find someone to proofread the paper before the submission. 3) Shorten the empirical study. 4) Need to fix ASE reviewers' questions about the evaluation, especially adding more details. 5) release the source code.}
Software maintenance activities are known to be generally expensive and challenging~\cite{planning2002economic} and one of the most important maintenance tasks is to handle bug reports~\cite{anvik2005coping}.
A good bug report is detailed with clear information about what happened and what the user expected to happen.
It goes on to contain a reproduction step or stack trace to assist developers in reproducing the bug, and supplement information such as screenshots, error logs, and environments.
As long as this bug report is accurate, it should be straightforward for developers to reproduce and fix.

Bugs are often encountered by non-technical users who will document a description of the bug with steps to reproduce it.
However, clear and concise bug reporting takes time, especially for non-developer or non-tester users who do not have that expertise and are not willing to spend that much effort~\cite{aranda2009secret, bettenburg2008extracting}.
A poorly written report would be even poorly interpreted~\cite{chaparro2019assessing,erfani2014works,bettenburg2008makes}, which often devolves into the familiar repetitive back and forth and the need to nag that comes with every single bug.
That effort may further prohibit users' contribution to bug reporting.
%bug reports differ in formats

Compared to writing it down with instructions on how to replicate, video-based bug reports significantly lower the bar for documenting the bug.
First, it is easy to record the screen as there are many tools available~\cite{web:BugClipper, web:TestFairy}, some of which are even embedded in the operating system by default like iOS~\cite{web:iosrecord} and Android~\cite{web:androidrecord}.
Second, video recording can include more detail and context such as configurations, and parameters, hence it bridges the understanding gap between users and developers.   
That convenience may even better engage users in actively providing feedback to improve the app.

Despite the pros of the video-based bug report, it still requires developers to manually check each frame in the video and repeat it in their environment.
According to our empirical study of 13,587 bug recordings from 647 Android apps in Section~\ref{sec:empirical}, one video is of 148.29 frames on average with a varied resolution for manual observation.
In addition, only 6.8\% of video recordings start from the app launch and most recordings begin 2-7 steps before the bug occurrence, indicating that developers need to guess steps to the entry frame of the video by themselves.
Therefore, it is necessary to develop an automated bug replay tool from video-based bug reports to save developers' effort in a bug fix.

There are many related works on bug replay but rarely related to visual bug reports.
Some researchers~\cite{zhao2019recdroid, fazzini2018automatically,zhao2019automatically,song2020bee,thummalapenta2012automating} leverage the natural language processing methods with program analysis to generate the test cases from the textual descriptions in bug reports.
However, those approaches do not apply to video-based bug reports.
There are platforms providing both video recording and replaying functionalities~\cite{lam2017record, qin2016mobiplay, bernal2020translating, web:androidstudio} which also store the low-level program execution information. %require source code
They require the framework installation or app instrumentation which is too heavy for end users.
Users tend to use general recording tools to get the video that only contains the visual information, according to our observation in Section~\ref{sec:observation}.
%That is also the reason why
% That also applies to crowd testing platform.
% \chen{Please briefly summerize related works and their limitation as our motivation or novelty.}
Automation for processing general recordings to reproduce bugs is necessary and would help developers shift their focus toward bug fixing.

To automate the analysis of bug recordings, we introduce \tool, a light-weight image-processing approach to automatically replay the video (GIF) based bug reports for Android apps.
First, we extract keyframes (i.e., fully rendered GUIs) of a recording by comparing the similarity of consecutive frames.
Second, a sequence of located keyframes is then mapped to the GUI states in the existing UTG (UI Transition Graph) of the app by calculating image similarity based on pixel and structural features. 
Third, given the mapped sequence, we propose a novel algorithm to not only address the defective mapped sequence, but also auto-complete the missing trace between app launch to the entry frame of the video, resulting in an optimal execution trace to automatically repeat the bug trigger.
% \chen{Please briefly introduce our approach, especially highlighting that our approach works in some very challenging situations mentioned above.}
% Extraction of keyframes of bug reports, mapping GUI to states in UTG, and generate the execution trace to automatically repeat the bug trigger. 

To evaluate the effectiveness of our tool, we create the groudtruth by manually labelling 61 video recordings from 31 apps.
As our approach consists of three main components, we evaluate them one by one including keyframe location, GUI mapping, and trace generation.
In all tasks, our approach significantly outperforms other baselines and successfully reproduce 82\% video recordings.
Apart from the accuracy of our tool, we also evaluate the usefulness of our tool by conducting a user study on replaying bugs from 10 real-world video recordings in GitHub.
Through the study, we provide the initial evidence of the usefulness of \tool for bootstrapping bug replay.

The contributions of this paper are as follows:
\begin{itemize}
	\item The first light-weight image-processing based approach, \tool, to reproduce bugs for Android apps directly from the GIF recordings with code released for public\footnote{ \url{https://github.com/sidongfeng/gifdroid}}.
    \item A motivational empirical study of large-scale recordings within real-world issue reports from GitHub including their content, length, etc.
	\item A comprehensive evaluation including automated experiments and a user study to demonstrate the accuracy, efficiency and usefulness of our approach. 

\end{itemize}

\section{Motivational Mining Study}
\label{sec:empirical}
While the main focus and contribution of this work is developing an approach to automatically replay the visual bug reports, we still carry out an empirical study to understand the characteristics of visual bug reports.
The overall status of visual bug reports can clearly frame the context and motivation of this work and their characteristics will be taken into consideration in our approach design.
But note that this motivational study just aims to provide an initial analysis towards developers supporting visual bug reports, and a more comprehensive empirical study would be needed to deeply understand it.
% \chen{Since most reviewers do not appreciate our empirical study, we should try our best to SHORTEN or even remove this section.}

\begin{comment}
To better understand the visual reports in real-world practice, we carry out an empirical study to explore what kinds of visual recordings exist.
In order to focus on helping developers to replay the bug scenario, we explore the general statistics of visual bug recordings, so as to motivate us to develop an automated approach to generate execution trace, given a visual recording.
\end{comment}

\begin{figure}  
	\centering 
	\includegraphics[width=0.95\linewidth]{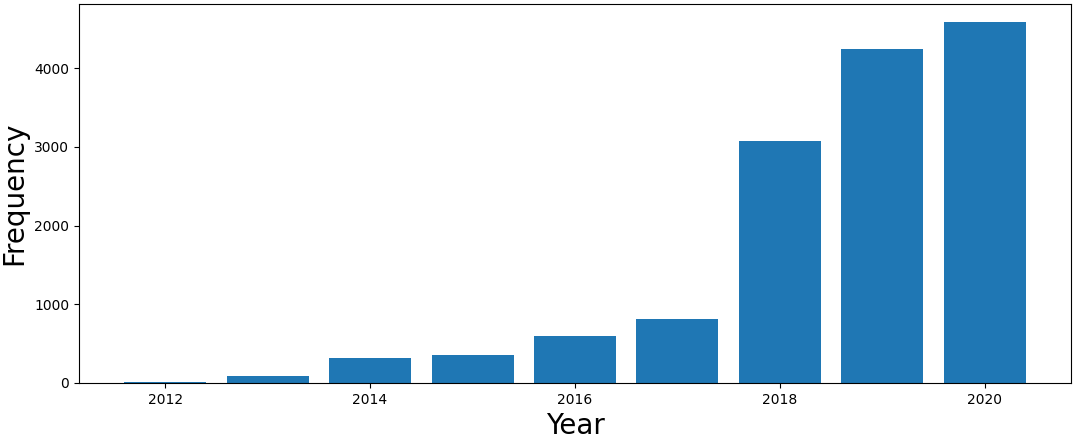}
	\caption{Number of recordings in GitHub from 2012 to 2020} 
	\label{fig:trend}
\end{figure}

\subsection{Data Collection}
We choose F-droid~\cite{web:fdroid} as the source of our study subjects, as it contains a large set of apps (1,274 at the time of our study) covering diverse categories such as connectivity, financial, multimedia.
All apps are open-source hosted on platforms like GitHub~\cite{web:github} which makes it possible for us to access their issue repositories for analysis.

We built a web crawler to automatically crawl the visual bug reports from issue repositories of these apps containing visual recordings (e.g., animation, video) with suffix names like .gif, .mp4, etc., or URL from video sharing platforms (e.g., GIPHY\footnote{GIPHY is the most popular animated GIF sharing platform, serving 700 million users}~\cite{web:giphy}, YouTube~\cite{web:youtube}).
%Note that we also crawl the video file from the URL.
It took us two weeks to scan 408,272 issues from 1,274 apps and finally mined 13,587 visual recordings from 7,230 issues (647 apps).
The dataset consists of 8,698 GIFs and 4,889 videos.
As shown in Figure~\ref{fig:trend}, attaching visual recording in the issue report is more and more popular.
%\sidong{A trend of using the visual recordings are shown in Fig.~\ref{fig:trend}.}
%There are more GIFs than videos due to the maximum size of the upload file i.e., 25MB.
Within the video, there is always some extra information (e.g., caption, whole screen of the PC, face of speaker) in addition to the app usage screen.
%And also compared to GIF, there may exist scene interference in video recordings, such as manual operation on the screen, I/O monitor, captioning, etc.
Since there are more GIFs than videos, we focus on the GIF visual recording for brevity in this paper.
% as it takes a large proportion of the dataset (64\%) and most of it is about exact app screen recording.

\begin{comment}
\begin{figure*}
	\centering
	\subfigure[Bug Replay]{
		\includegraphics[width = 0.31\linewidth]{figures/category1.png}
		\label{fig:category1}}
	\hfill
	\subfigure[Feature Request]{
		\includegraphics[width = 0.31\linewidth]{figures/category2.png}
		\label{fig:category2}}		
	\hfill
	\subfigure[Issue Fixed]{
		\includegraphics[width = 0.31\linewidth]{figures/category3.png}
		\label{fig:category3}}
	\caption{Examples of three kinds of recordings in issue reports. \chen{1)Just use 1/2 example for the first category 2) make it single column 3) highlight the words in the description 4)the first example is not good, and please replace with another one.}}
	\label{fig:category}
\end{figure*}
\end{comment}

\begin{figure}
	\centering 
	\includegraphics[width=0.95\linewidth]{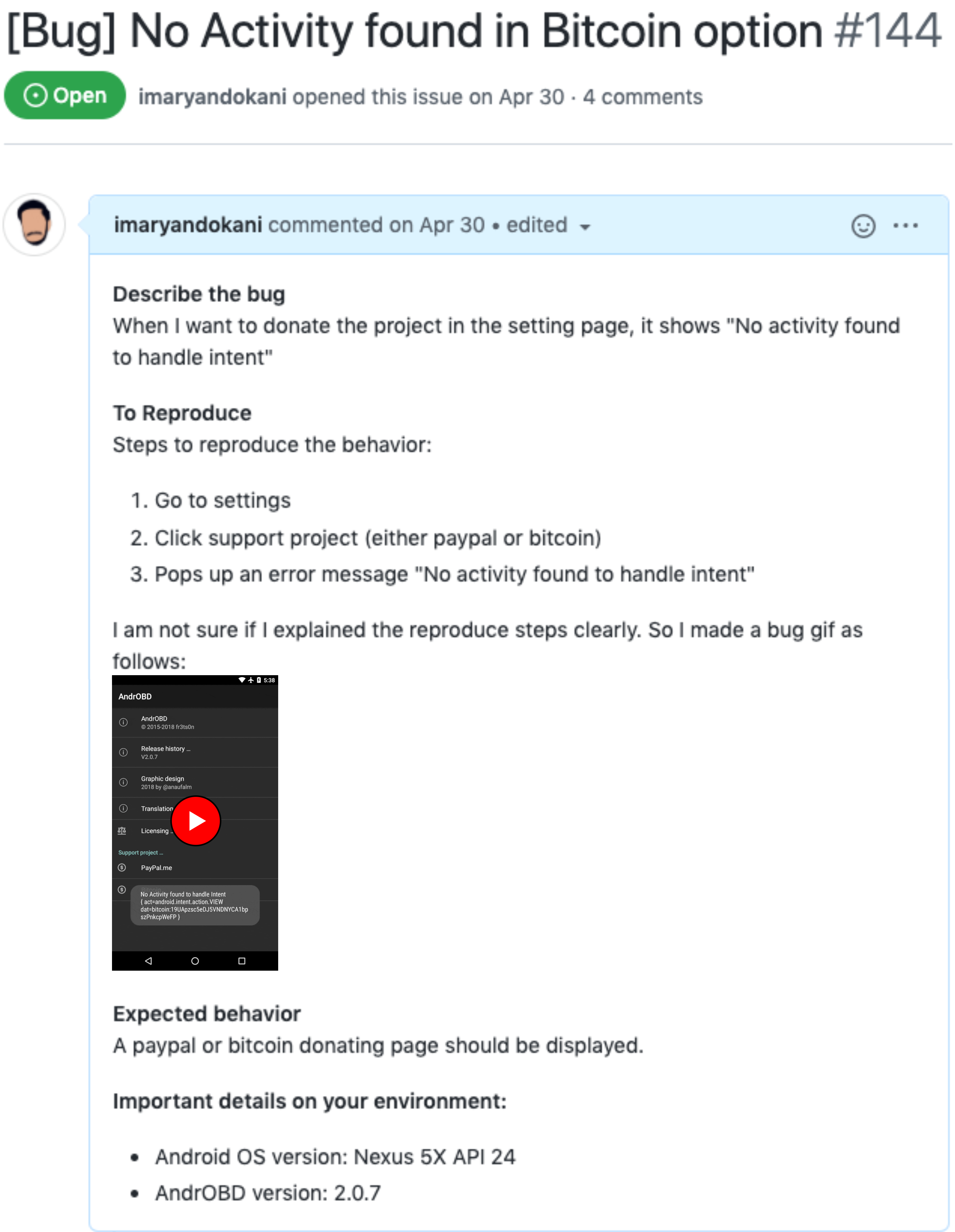}
	\caption{Example of the issue reports for Bug Replay. The reporters provide a visual recording (i.e., gif) to help developers understand the bug report more clearly.} 
	\label{fig:bugExample}
	\vspace{3pt}
\end{figure}

\subsection{What are in Visual Recordings?}
\label{sec:observation}
To understand the content within these visual recordings, the first two authors manually check 1,000 (11.5\%) GIFs and their corresponding descriptions in the issue report.
\begin{comment}
Given those visual recordings, the first two authors individually check 1,000 (11.5\%) manually with also its corresponding description in the issues.
During the manual examination process, we notice that those visual recordings are highly variable, and the understanding would facilitate the design and evaluation of our approach.
We elaborate our findings of visual recordings in two parts: visual recording category, and visual recording creation.
\textbf{visual recording category:}
\label{sec:gif_category}
Given the issue descriptions, the scenario depicted in the recording varies.
\end{comment}
Following the Card Sorting~\cite{spencer2009card} method, we classify those GIFs into three categories:

(i) Bug Replay (60\%).
%As the users only have functional knowledge of the app, whereas the developers working on an app tend to have intimate code level knowledge.
%While users submit bugs reports in the most amount of detailed he could, including the steps to reproduce the bug, the expected and actual behaviour,and the testing environment, it is still hard for developers to understand the bugs.
Most screen recordings are about the bug replay including detailed procedures in triggering the bugs, especially for those requiring complicated actions with one example seen in Figure~\ref{fig:bugExample}\footnote{\url{https://github.com/fr3ts0n/AndrOBD/issues/144}}.
%To clarify the bug report, users was asked to upload a visual recording to show the bug intuitively as shown in .
The visual recordings bridge the lexical gap between users and developers and help developers read and comprehend the bug report reasoning about the problem from the high-level description.

(ii) Issue Fixed (17\%).
Once the issue is fixed, developers tend to record the screen to display the update changes, along with a checklist for project owners to approve the pull request, 
%as shown in Fig.~\ref{fig:category3}\footnote{https://github.com/fossasia/neurolab-android/pull/555}.
Also, those recordings are used for app introduction or answering users' feature requests.

(iii) Feature Request (23\%).
Users may use recordings to ask for missing functionality (e.g., provided by other apps) or missing content (e.g., in catalogues and games), and sharing ideas on how to improve the app in future releases by adding or changing features.
%For example, in Fig.~\ref{fig:category2}\footnote{https://github.com/mozilla-mobile/fenix/issues/16004}, the user attached a visual recording that displays the demand feature in other apps, providing a visually appearance of the feature.  
%\chen{In the future, we can do some research in this part as it is highly related crowd wisdom which is especially interesting in HCI domain.}

\begin{comment}

\subsection{How do Users Create Recordings?}
%\textbf{visual recording creation:}
By analyzing these recordings and checking materials online, we find that there are three commonly-used practices to create a GIF recording including video conversion, on-device recording, and emulator recording.
First, due to the size limit, some users convert the video to GIF, resulting in a quality degradation including resolution, frame number, etc.
Second, some users directly record the app usage in their device and save it as GIF using apps like GIPHY Cam~\cite{web:giphycam} or GIF Maker~\cite{web:gifmaker}. 
This kind of GIFs usually has a watermark.
Third, developers directly record it in the emulator screen running on the desktop.
Thus, those GIFs normally contain a mobile template border.
As there are multiple ways to create a visual recording, prior researches~\cite{bernal2020translating} that heavily rely on specific recording constraints cannot be generalized.	
\end{comment}

\begin{table}
\centering
\caption{Recording resolution.}
\label{tab:resolution}
\begin{tabular}{ccc} 
        \hline
        \bf{Ratios} & \bf{Occurrences} & \bf{Resolution} \\ 
        \hline
        16:9 & 288 (48\%) & \makecell{1920$\times$1080 \\ 1280$\times$720 \\ 960$\times$540} \\
        \hline
        5:3 & 63 (10.5\%) & \makecell{1280$\times$800 \\ 800$\times$480} \\
        \hline
        3:2 & 21 (3.5\%)  & \makecell{900$\times$600 \\ 480$\times$320}\\
        \hline
    \end{tabular}
\vspace{0.4cm}
\end{table}

\begin{figure}
	\centering
	\includegraphics[width=0.95\linewidth]{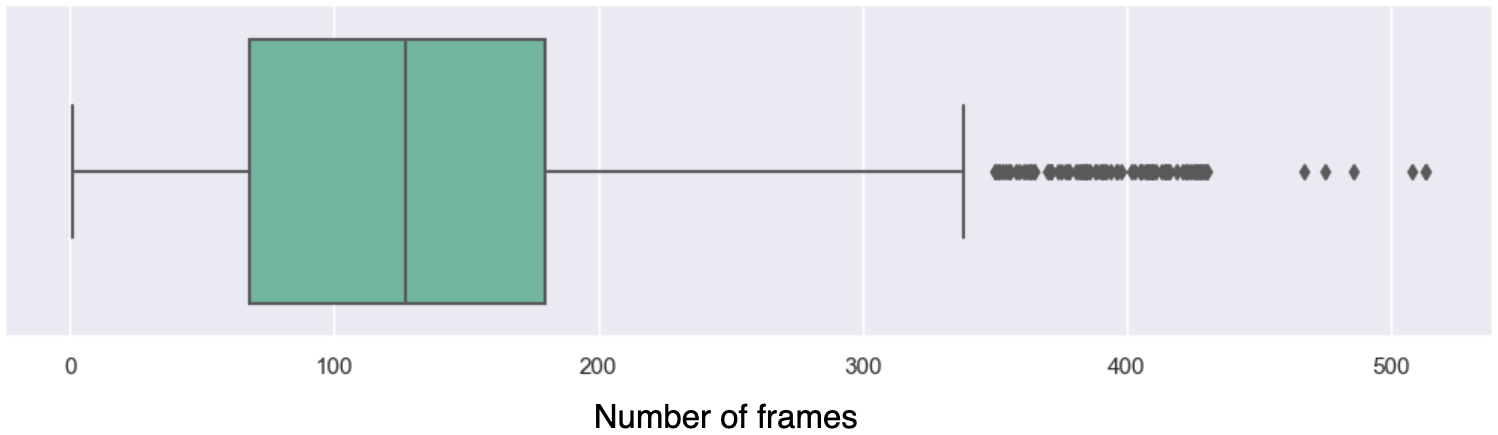}
	\caption{Number of frames of each recording.}
	\label{fig:frames}
	\vspace{0.2cm}
\end{figure}

% \begin{minipage}[c]{0.5\linewidth}
%     \vspace{0.8cm}
%     \centering
%     \small
%     \tabcolsep=0.07cm
%     \begin{tabular}{ccc} 
%         \hline
%         \bf{Ratios} & \bf{Occurrences} & \bf{Resolution} \\ 
%         \hline
%         16:9 & 288 (48\%) & \makecell{1920$\times$1080 \\ 1280$\times$720 \\ 960$\times$540} \\
%         \hline
%         5:3 & 63 (10.5\%) & \makecell{1280$\times$800 \\ 800$\times$480} \\
%         \hline
%         3:2 & 21 (3.5\%)  & \makecell{900$\times$600 \\ 480$\times$320}\\
%         \hline
%     \end{tabular}
%     \captionof{table}{Recording resolution}
%     \label{tab:resolution}
% \end{minipage}
% \hfill
% \begin{minipage}[c]{0.4\linewidth}
%     \centering
%     \includegraphics[width=0.85\linewidth]{figures/boxplot.png}
%     \captionof{figure}{No. of Frames. \chen{Since e got enough space, we can separate the table and figure. If so, please draw the box plot horizontally.}}
%     \label{fig:frames}
% \end{minipage}

%\subsection{General Statistics of Visual Bug Recording}
\subsection{What are Characteristics of Bug Recordings?}
\label{sec:general_statistics}
Since the target of this work is to auto-replay the bug recording, we further analyze the characteristics of 600 bug recordings classified in Section~\ref{sec:observation}.
Due to the difference of devices or users' resize of the video, there may be different resolutions and aspect ratios as summarized in Table~\ref{tab:resolution}.
About half of bug recordings are of 16:9 aspect ratio with varied resolution from high-quality 1920x1080 to 960x540.
There are also some minor aspect ratios and resolutions as users may post-process the recording by resizing or cropping arbitrarily, according to our observation. 

Figure~\ref{fig:frames} depicts the number of frames in visual bug recording.
On average, there are 148.29 frames per video and some of them are even with more than 500 frames which require developers to manually check and replay in their own settings.
%(e.g., 12.86 seconds).
We also find that only 41 recordings (6.8\%) start from the launch of the app, while most recordings start from 2-7 steps before the bug occurrence. 
It indicates that there is no explicit hint to guide developers to the entry frame of the recording and that barrier may negatively influence developers' efficiency.
As the important information to show users' actions, some recordings contain touch indicators (e.g., circle or arrow of the touch position) which is an important resource to replay the bug~\cite{bernal2020translating}.
However, there are only 155 recordings (25.8\%) enabling the ``Show Touches'' option in our dataset.
Developers especially novice ones have to guess and try the potential actions to trigger the target page from the current page.

\begin{comment}
Since we only focus on the first category of visual recordings i.e., bug replay, we analyze the general statistics of 600 visual recordings classified in Section~\ref{sec:gif_category}.
Users may resize or crop the visual recording before the submission, therefore, we adopt aspect ratios for resolution statistics, which is more intuitive.
Table~\ref{tab:resolution} shows the statistics for the top-3 aspect ratios with their occurrences and resolutions that occur frequently.
The resolutions are consistent with the common mobile screen resolution~\cite{web:mobile}. 
\sidong{Figure}~\ref{fig:frames} depicts the number of frames in visual bug recording, on average 148.29 frames (e.g., 12.86 seconds).
However, those recordings contain relatively few GUI transitions (e.g., an average of 4.68) due to the time cost on rendering and user transiting.
Besides, only 41 recordings (6.8\%) record from the launch of the app.
Analyzing touch indicator is a common method to reproduce the scenario from visual recording~\cite{bernal2020translating,bao2017extracting}. 
There are only 155 recordings (25.8\%) enabled the "Show Touches" option.
It does make sense due to the issue repository is serving for both technical and non-technical users, unless the "Show Touches" option in the "Developer Options" settings is enabled deliberately, the recording will not have a touch indicator.
\end{comment}

\vspace{0.3cm}

\noindent\fbox{
	\parbox{0.95\linewidth}{
		\textbf{Summary}: By analyzing issue reports from 1,274 existing apps crawled from F-Droid, 60\% of them are with recordings for bug replay.
		A large number of frames and varied resolution make it difficult for developers to replay them in their setting. 
		Such phenomenon is further exacerbated as 74.2\% of recordings are without touch indicators on the screen.
		These findings confirm the necessity and difficulty of the replay of visual bug reports, and motivate our approach development for automatic replaying the recordings for developers and testers.
	}
}

\section{Approach}
\begin{figure*}[h!]
	\centering 
	\includegraphics[width=0.85\textwidth]{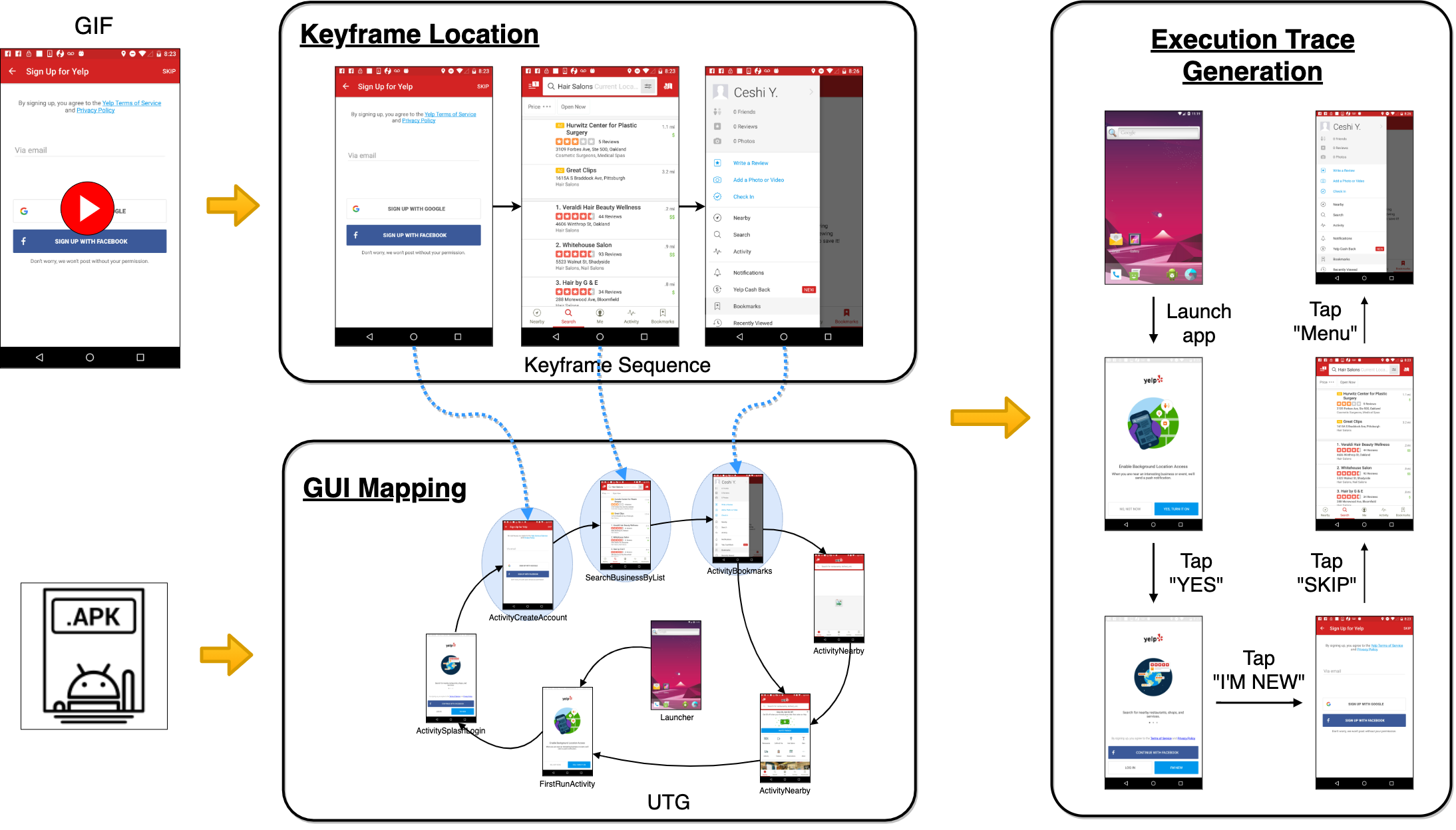}  
	\caption{The overview of \tool.} 
	\label{fig:approach}
	\vspace{0.3cm}
\end{figure*}
%Inspired by the GUI exploration tool that can automatically produce an GUI Transition Graph,
Given an input bug recording, we propose an automated approach to localize a sequence of keyframes in the GIF and subsequently map them to the existing UTG (UI Transition Graph) to extract the execution trace.
The overview of our approach is shown in Figure~\ref{fig:approach}, which is divided into three main phases: 
(i) the \textit{Keyframe Location} phase, which identifies a sequence of keyframes of an input visual recording,
(ii) the \textit{GUI Mapping} phase that maps each located keyframe to the GUIs in UTG, yielding an index sequence,
and (iii) the \textit{Execution Trace Generation} phase that utilizes the index sequence to detect an optimal replayable execution trace.

\subsection{Keyframe Location}
\label{ref:phase1}
Note that GUI rendering takes time, hence many frames in the visual recording are showing the partial rendering process.
The goal of this phase is to locate keyframes i.e., states in which GUI are fully rendered in a given visual recording.
%A sequence of keyframes offers a concise representation of the visual bug recording by showing the states between steady frames.

\subsubsection{Consecutive Frame Comparison}
\label{ref:frame_comparison}
Inspired by signal processing, we leverage the image processing techniques to build a perceptual similarity score for consecutive frame comparison based on Y-Difference (or Y-Diff).
YUV is a color space usually used in video encoding, enabling transmission errors or compression artifacts to be more efficiently masked by the human perception than using a RGB-representation~\cite{chen2009compression,sudhir2011efficient}.
Y-Diff is the difference in Y (luminance) values of two images in the YUV color space. 
We adopt the luminance component as the human perception (a.k.a human vision) is sensitive to brightness changes.
In the human visual system, a majority of visual information is conveyed by patterns of contrasts from its brightness changes~\cite{zeki1993vision}.
Furthermore, luminance is a major input for the human perception of motion~\cite{livingstone2002vision}. 
Since people perceive a sequence of graphics changes as a motion, consecutive images are perceptually similar if people do not recognize any motions from the image frames.
%Therefore, we leverage Y-Diff as the perceptual similarity measurement for consecutive frame comparison.
%\chen{May simplify this section if more space is needed.}

Consider a visual recording $\big\{ f_{0}, f_{1}, .., f_{N-1}, f_{N} \big\}$ , where $f_{N}$ is the current frame and $f_{N-1}$ is the previous frame.
To calculate the Y-Diff of the current frame $f_{N}$ with the previous $f_{N-1}$, we first obtain the luminance mask $Y_{N-1}, Y_{N}$ by splitting the YUV color space converted by the RGB color space.
Then, we apply the perceptual comparison metric, SSIM (Structural Similarity Index)~\cite{wang2004image}, to produce a per-pixel similarity value related to the local difference in the average value, the variance, and the correlation of luminances.
In detail, the SSIM similarity for two luminance masks is defined as: 
\begin{equation}
	SSIM(x,y) = \frac{(2\mu_x\mu_y + C_1) + (2 \sigma _{xy} + C_2)}	{(\mu_x^2 + \mu_y^2+C_1) (\sigma_x^2 + \sigma_y^2+C_2)} 
	\label{eq:SSIM}
\end{equation}
	
where \(x, y\) denote the luminance masks \( Y_{N-1}, Y_{N}\), and \(\mu_x\), \(\sigma_x\), \(\sigma_{xy}\) are the mean, standard deviation, and cross correlation between the images, respectively.
\(C_{1}\) and \(C_{2}\) are used to avoid instability when the means and variances are close to zero. 
A SSIM score is a number between 0 and 1, and a higher value indicates a strong level of similarity.

\begin{figure}  
	\centering 
	\includegraphics[width=0.95\linewidth]{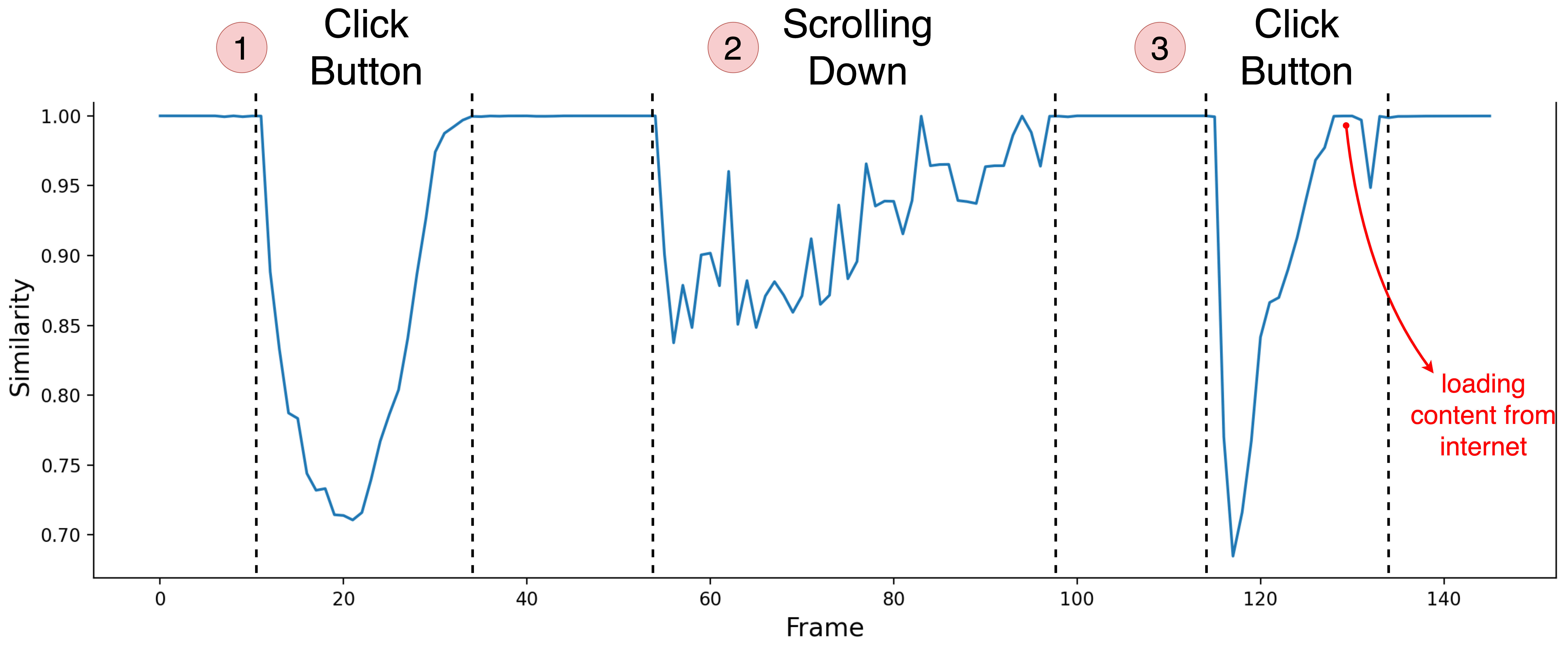}
	\caption{An illustration of the Y-Diff similarity scores of consecutive frames in the visual recording.} 
	\label{fig:timeframe}
	\vspace{0.1cm}
\end{figure}

\subsubsection{Keyframe Identification}
To make decisions on whether the frame is a keyframe, we look into the similarity scores of consecutive frames in the visual recording as shown in Figure~\ref{fig:timeframe}.
The first step is to group frames belonging to the same atomic activity according to a tailored pattern analysis.
This procedure is necessary because discrete activities performed on the screen will persist across several frames, and thus, need to be grouped and segmented accordingly.
There are 3 types of patterns, i.e., \textit{instantaneous transitions}, \textit{animation transitions}, and \textit{steady}.

\textit{(1) instantaneous transitions}: As shown in Figure~\ref{fig:timeframe} Activity 1 (clicking a button), the similarity score starts to drop drastically which reveals an instantaneous transition from one screen to another.
In addition, one common case is that the similarity score becomes steady for a small period of time $t_{s}$ between two drastically droppings as shown in Figure~\ref{fig:timeframe} Activity 3. 
The occurrence of this short steady duration $t_{s}$ is because GUI has not finished loading.
While the GUI layout of GUI rendering is fast, resources loading may take time. 
For example, rendering images from the web depends on device bandwidth, image loading efficiency, etc.

% First, GUI has not finished loading. 
% While the GUI layout of GUI rendering is fast, resources loading may take time. 
% For example, rendering image from web depends on device bandwidth, image loading efficiency, etc.
% Second, the refresh rate of the screen may be too fast.
% The screen will render the same GUI until the device finishes processing~\cite{web:highfps}.
% \chen{Do not understand the second reason.}

\textit{(2) animation transitions}:
Figure~\ref{fig:timeframe} Activity 2 shows the similarity sequence of the transitions where animation effects are used, for example, the "scrolling" event.
That is, over a period of time, the similarity score continues to increase slightly.

\textit{(3) steady}:
Figure~\ref{fig:timeframe} gives an example of a GUI steady state where the consecutive frames are similar for a relatively long duration.
We have empirically set 0.95\footnote{We set up that value by a small-scale pilot study.} as the threshold to decide whether two frames are similar, and 5 frames as the threshold to indicate a steady state.

%We select the last frame in the GUI steady state group as a keyframe, denoting the fully rendered GUI. 
%\sidong{Please confirm.}

% Since user always initiate the next interaction after the GUI is completely rendered.
% To this perception, we identify the last frame in the GUI steady state group as the keyframe.

% Measuring similarity score over time reveals that the similarity is higher and stable between frames within a key snippet than instantaneous transitions from one screen to another (e.g., click) and during transitions where digital animation effects used (e.g., scroll, pop-up).

% digital variants of YUV color spaces
% These colour spaces separate RGB into luminance and chrominance information and are useful in compression applications~\cite{ford1998colour}

\subsection{GUI Mapping}
\label{ref:phase2}
It is easy for developers to get the UTG of their own app~\cite{chen2019storydroid,yang2018static}.
Therefore, instead of inferring actions from the recording~\cite{zhao2019actionnet, bernal2020translating}, we directly map keyframes extracted from the recording to states/GUIs within the UTG.
%The goal of this phase is to convert the keyframe sequence into an index sequence by using image searching techniques to map each keyframe to the GUI screenshots in the UTG.
To achieve this, we compute the image similarity between the keyframe and each GUI screenshot based on both pixel and structural features.
The GUI screenshot that has the highest similarity score is regarded as the index of the keyframe.

\begin{figure}
	\centering
	\includegraphics[width=0.85\linewidth]{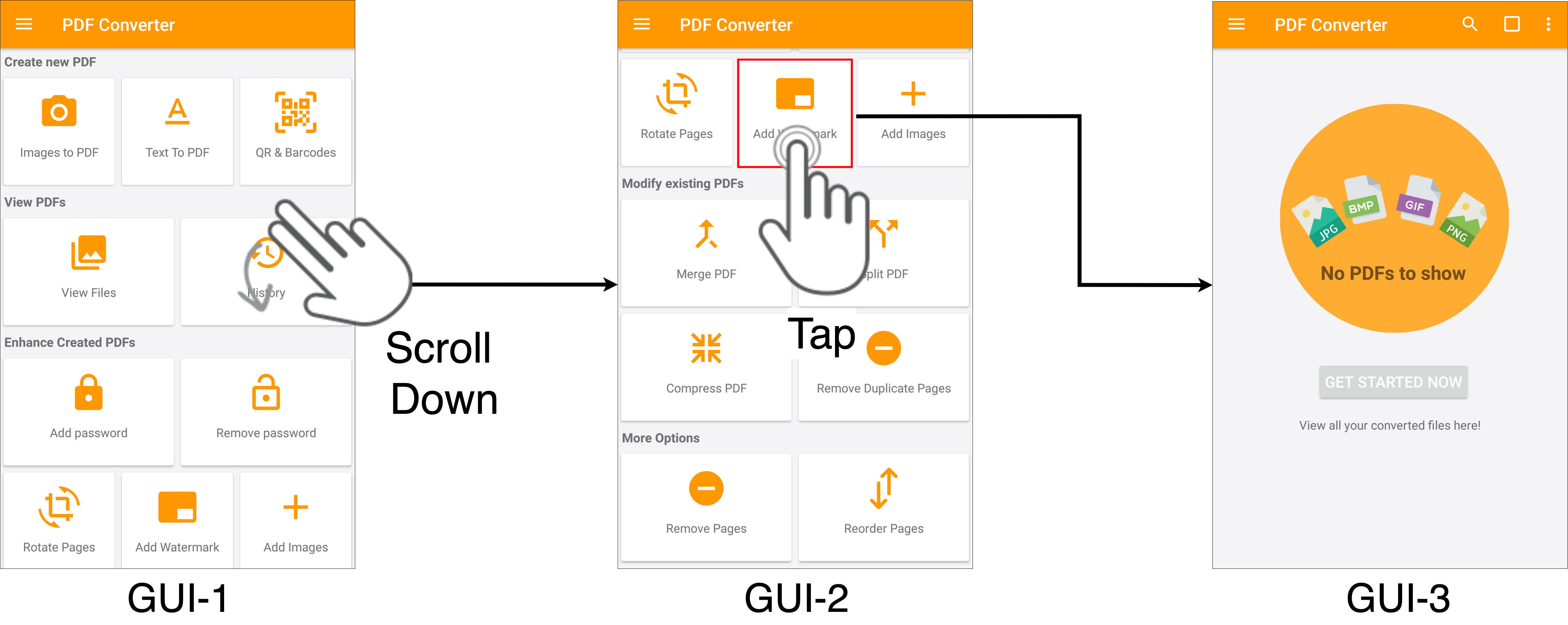}
	\caption{An example of partial UTG on mobile app GUI proceeding from state GUI-1 to state GUI-3.}
	\label{fig:utg}
	\vspace{0.1cm}
\end{figure}
% \chen{The last two states are the same, any way to show different ones?}

\subsubsection{UTG Construction}
\label{sec:utg_collection}
A GUI transitions graph (UTG) of an Android app is  widely used to illustrate the transitions across different GUIs triggered by typical elements such as toasts (pop-ups), text boxes, text view objects, spinners, list items, progress bars, checkboxes. 
In Figure~\ref{fig:utg} we illustrate how a UTG emerges as a result of user interactions in an app.
% \chen{May simplify if more space is needed.}
%  (i.e., tapping and editing textview) 
Starting with the GUI-1, the user ``scrolls down'' to the bottom of the page (GUI-2). 
When the user ``taps'' a button, the GUI transits to the GUI-3.
There are many tools to construct a UTG, either manually~\cite{web:sketch} or automatically~\cite{li2017droidbot,azim2013targeted}.
In this paper, we adopt the Firebase~\cite{web:firebase}, a widely-used automated GUI exploration tool developed by Google, while other tools can also be used.

\subsubsection{Feature Extraction}
The basic need for any image searching techniques is to detect and construct a discriminative feature representation~\cite{deka2017rico,deka2016erica}.
A proper construction of features can improve the performance of the image searching method~\cite{li2021screen2vec}.
According to our observation, in addition to the pixel features as that in natural images, GUI screenshots are also of additional structural features i.e., the layout of different components in one page. 
Therefore, we adopt a hybrid method based on two types of features, SSIM (Structural Similarity Index)~\cite{wang2004image} and ORB (Oriented FAST and Rotated BRIEF)~\cite{rublee2011orb}, for searching the mapping GUI screenshots in the UTG for the keyframe.

While SSIM detects the features within pixels and structures (i.e., a detailed description is demonstrated in Section~\ref{ref:frame_comparison}), it still has several fundamental limitations that exist in visual recordings, e.g., image distortion~\cite{wang2011ssim,lee2016improved}.
To address this, we further supplement a local invariant feature extraction method, ORB.
Given an image, ORB first detects the interest points, indicating at which the direction of the boundary of the object changes abruptly or intersection point between two or more edge segments.
% To address the scaling invariant problem, ORB uses a multiscale image pyramid, which consists a downsampled sequences of images at different resolutions, resulting in a local invariant interest points. 
Then, ORB converts each interest point into an n-bits binary feature descriptor, which acts as a ``fingerprint'' that can be used to differentiate one feature from another. 
A feature descriptor of an interest point is computed by an intensity difference test $\tau$:
\begin{equation}
	\tau(p; x,y) = 
	\begin{cases}
		1 & \text{if p(x) $<$ p(y)}\\
		0 & \text{otherwise}
	\end{cases}
	\label{eq:orb_descriptor}
\end{equation}
where $p(x), p(y)$ are intensity values at pixel $x, y$ around the interest point. 
Due to the characteristic of local feature extraction, ORB features remain invariant of scale, brightness, and also maintain a certain stability of affine transformation, and noise.
% , that is if the first pixel is brighter than the second, it assigns the corresponding bits to 1 else 0.

\subsubsection{Similarity Computation}

Based on the features extracted by SSIM and ORB, we compute a similarity value $S_{ssim}$ and $S_{orb}$, respectively.
To compute $S_{orb}$, we adopt the Brute Force algorithm~\cite{rublee2011orb}, which compares the hamming distance between feature descriptors. 
We compute the $S_{ssim}$ using Equation~\ref{eq:SSIM} based on similarity on luminance, contrast, and structure.
We then further determine the similarity $S_{comb}$ between the keyframe and states in UTG by combining two feature similarities score:
%Then, we use a linear combination approach to produce an aggregate similarity value:
\begin{equation}
	S_{comb} = w \times S_{orb} + (1 - w) \times S_{ssim}
	\label{eq:sim_comb}
\end{equation}
where $w$ is a weight for $S_{ssim}$ and $S_{orb}$, taking a value between 0 to 1.
Smaller $w$ value weights $S_{ssim}$ more heavily, and larger value weights $S_{orb}$ more heavily.
We empirically choose 0.5 as the $w$ value for the best performance. 

Based on the combined similarity between the keyframe and each GUI screenshot, we select the highest score to be the index of the keyframe.
Consequently, a sequence of keyframes is converted to a sequence of the index in the UTG.

\begin{figure}  
	\centering 
	\includegraphics[width=0.95\linewidth]{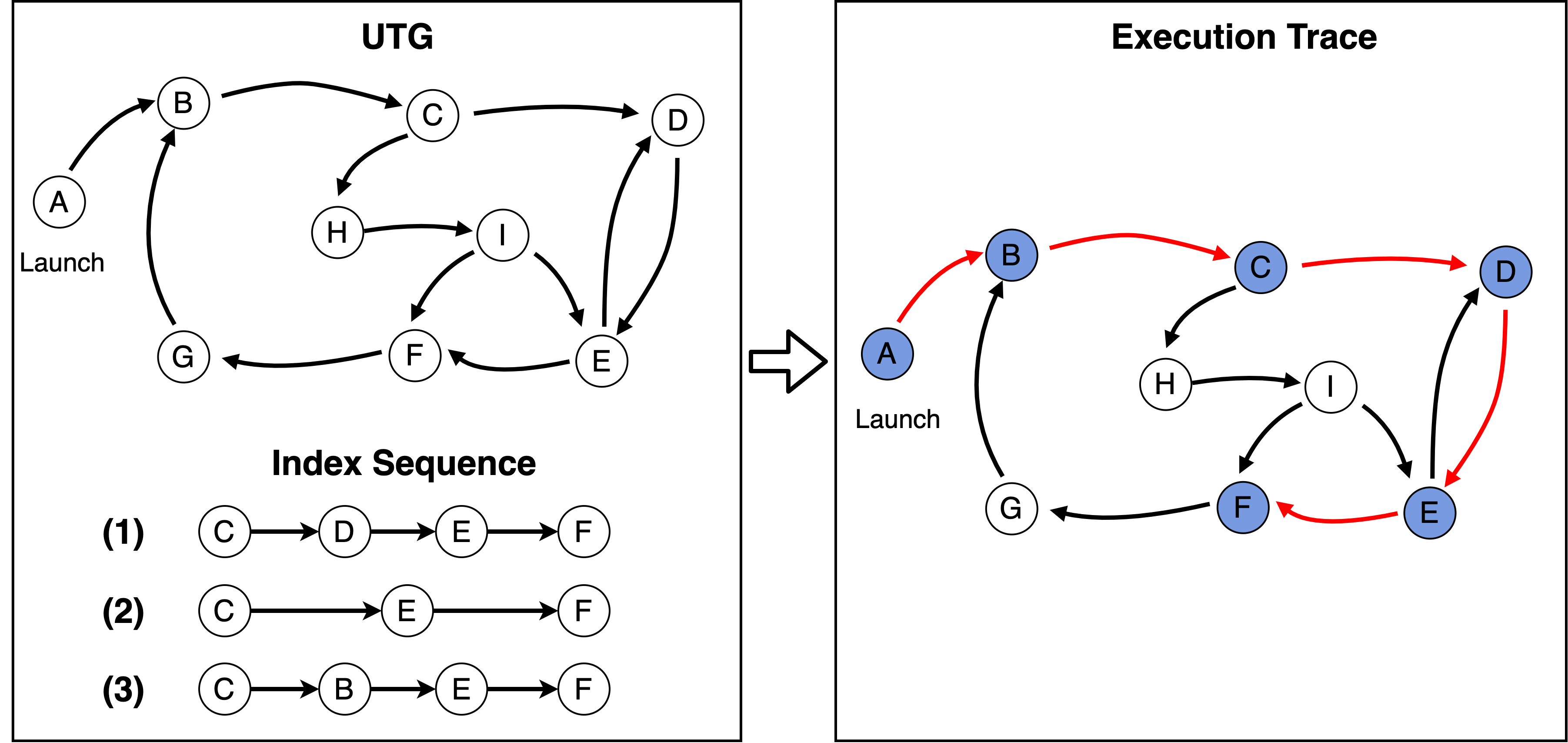}  
	\caption{Illustration of the execution trace generation. Index sequence 2,3 indicate two types of defective sequences, i.e., missing $\{D\}$ and wrong mapping to $\{B\}$, respectively. } 
	\label{fig:lcs}
\end{figure}

\subsection{Execution Trace Generation}
\label{sec:executiontrace}
After mapping keyframes to the GUIs in the UTG, we need to go one step further to connect these GUIs/states into a trace to replay the bug.
However, this process is challenging due to two reasons.
First, the extracted keyframe (Section~\ref{ref:phase1}) and mapped GUIs (Section~\ref{ref:phase2}) may not be 100\% accurate, resulting in a mismatch of the groundtruth trace.
For example in Figure~\ref{fig:lcs}, $\{D\}$ is missed in the index sequence 2 and the second keyframe is wrongly mapping to $\{B\}$ in the index sequence 3.
Second, different from the uploaded GIF which may start the recording anytime, the recovered trace in our case must begin from the launch of the app. 

Therefore, the trace generation algorithm needs to consider both the wrong extraction/mapping in our previous steps, and the missing trace between the app launch and first keyframe in the visual bug report.
To overcome these issues, our approach first generates all candidate sequences in UTG between the app launch to the last keyframe from GIF.
By regarding the extracted keyframes as a sequence, our approach then further extracts the Longest Common Subsequence (LCS) between it and all candidate sequences.

\begin{algorithm}[t]
    \linespread{1.2}
    % \setstretch{1.4}
	\SetAlgoLined
	\SetNoFillComment
	\SetKwInOut{Input}{Inputs}
	\SetKwInOut{Output}{Output}
	\SetKwProg{DFS}{DFS}{}{}
	\SetKwProg{FindLCS}{FindLCS}{}{}
	\Input{$X$: index sequence $\{x_1, x_2, x_3, ..., x_n\}$; \\ 
		$G$: UTG graph; \\
		$s$: starting node in UTG;}
	\Output{execution trace}
	\tcc{Find the candidate sequences ($SEQs$) from app launch node to last index node}
	$SEQs \gets$ [] \;
	$SEQ \gets$ [] \;
	$visited \gets$ false for all nodes in UTG\;
	\DFS{$(s, x_n, visited, SEQ)$}{
		visited[$s$] $\gets$ true\ \tcp*{prevent acyclic sequence}
		$SEQ.append(s)$ \;
		\uIf{$s$ == $x_n$}{
			$SEQs.append(SEQ)$ \;
		}
		\Else{
			\ForEach{$v$ $\in$ $G$.Adj[$s$]}{
				\uIf{visited[$v$] == false}{
					\DFS{$(v, x_n, visied, SEQ)$}{}
				}
			}
		}
		\tcc{Backtrack to take another path}
		$SEQ.pop()$ \;
		visited[$s$] $\gets$ false \;
	}
	\tcc{Find the $LCSs$ between index sequence and each candidate sequence}
	\ForEach{$SEQ$ $\in$ $SEQs$}{
		let $com[0..m, 0..n]$ be new table \;
		initialize $com \gets 0$ \tcp*{prevent NULL sequence}
		\For{$i\gets1$ \KwTo $X.length$}{
			\For{$j\gets1$ \KwTo $SEQ.length$}{
				\uIf{$x[i] == SEQ[j]$}{
					$com[i,j] \gets com[i-1,j-1]+1$ \\
					%  $b[i,j] \gets "\nwarrow"$
				}
				\uElseIf{$com[i-1,j] \geq com[i,j-1]$}{
					$com[i,j] \gets com[i-1,j]$ \\
					% $b[i,j] \gets "\uparrow"$
				}
				\Else{
					$com[i,j] \gets com[i,j-1]$ \\
					% $b[i,j] \gets "\leftarrow"$
				}
			}
		}
		$LCSs \gets$ save ($FindLCS(com)$)
	}
	\tcc{Find the execution trace with longest LCSs and shortest sequences}
	$Trace \gets$ $SEQs$ [$max(LCSs)$ $\&$ $min(SEQs)$] \;
	return $Trace$
	\caption{Execution Trace Generation}
	\label{algorithm:lcs}
\end{algorithm}

The overflow of our approach can be seen in Algorithm~\ref{algorithm:lcs}.
Given an index sequence $X = \{x_1, x_2, ..., x_n\}$ (extracted by keyframe mapping) where $x_n$ is the last node, a UTG graph $G$, and an app launch node $s$.
To find acyclic paths (avoiding dead loops) from app launch node ($s$) to last index node ($x_n$), we adopt Depth-First Search traversal, that takes a path on $G$ and starts walking on it and check if it reaches the destination then count the path and backtrack to take another path. 
% If the path does not lead to the destination, discard the path. 
To avoid cyclic path, we record all visited nodes (Line 5), so that one node cannot be visited twice.
The output of the traversal is a set of acyclic path $SEQ = \{y_1, y_2, ..., y_m\}$ where $y_1 = s$ and $y_m = x_n$.
%\chen{Please add more line annotation when describing the algorithm}
We regard those acyclic paths as the candidate sequences for execution.
For example, in Figure~\ref{fig:lcs}, there are three candidate sequences from launch node $\{A\}$ to last index node $\{F\}$ are found (1)$\{A,B,C,D,E,F\}$, (2)$\{A,B,C,H,I,F\}$, (3)$\{A,B,C,H,I,E,F\}$.
Note that $\{A,B,C,D,E,D,E,F\}$ is omitted due to cyclic.

Next, for each candidate sequence $SEQ$, we adopt the dynamic programming algorithm to find the LCS to the index sequence $X$ (Lines 17-32), in respect to detect how many index nodes are covered in this candidate sequence.
The recursive solution to detect the LCS for each candidate sequence $SEQ$ to index sequence $X$ can be defined as:
\begin{equation}
	com[i,j] = 
	\begin{cases}
		$0$ & \text{if $i=0$ or $j=0$} \\
		$com[i-1,j-1]+1$ & \text{if $x[i]=seq[j]$} \\
		$max(com[i,j-1],com[i-1,j])$ & \text{if $x[i] \neq seq[j]$}
	\end{cases}
	\label{eq:lcs}
\end{equation}
where $com$ to be a table for $X$ and $SEQ$ (e.g., $com[i,j]$ is the relation between $X_i$ and $SEQ_j$).
A LCS can be found by tracing the table $com$.
We omit the details of tracing the table as $FindLCS(com)$ due to space limitations.
Note that we initialize first row and column of $com$ as zero to prevent the occurrence of NULL (Line 19).
For example, in Figure~\ref{fig:lcs}, the LCSs between index sequence and three candidate sequences are (1)$\{C,E,F\}$ (2)$\{C,F\}$ (3)$\{C,E,F\}$, respectively.

Once the LCSs are detected, we select the candidate sequence that has the longest LCS as the execution trace due to it replays most keyframes (or index nodes) in the visual recording.
Besides, our goal is to help developers reproduce the bug with the least amount of time/steps.
Therefore, we choose the optimal execution trace with the shortest sequence.
For example, in Figure~\ref{fig:lcs}, 
(2)$\{A,B,C,H,I,F\}$ is omitted due to it does not replay the most index nodes, (i.e., its LCS $\{C,F\}$ is not the longest).
(3)$\{A,B,C,H,I,E,F\}$ is omitted due to it is not the optimal trace (i.e., not the shortest sequence).
Therefore, the optimal execution trace (1)$\{A,B,C,D,E,F\}$ is generated based on the index sequence, even defective.

% Optimization strategies: 
% (1) Check if the frame is available in the firebase, as it could be the irrelevant state such as the mobile interface, Crash statement interface
% (2) Looping state

\begin{comment}
\begin{algorithm}[]
\SetAlgoLined
\SetKwProg{LCS}{LCS}{}{}
Detect LCSs between \textit{X} and each path \textit{Y}, and select the longest one \;

let $b[1..m, 1..n]$ and $c[0..m, 0..n]$ be new tables \;
\For{$i\gets1$ \KwTo $m$}{
$c[i,0] \gets 0$
}
\For{$j\gets0$ \KwTo $n$}{
$c[0,j] \gets 0$
}
\For{$i\gets1$ \KwTo $m$}{
\For{$j\gets1$ \KwTo $n$}{
\uIf{$x_i == y_j$}{
$c[i,j] \gets c[i-1,j-1]+1$ \\
$b[i,j] \gets "\nwarrow"$
}
\uElseIf{$c[i-1,j] \geq c[i,j-1]$}{
$c[i,j] \gets c[i-1,j]$ \\
$b[i,j] \gets "\uparrow"$
}
\Else{
$c[i,j] \gets c[i,j-1]$ \\
$b[i,j] \gets "\leftarrow"$
}
}
}

\caption{Longest Common Subsequence}
\end{algorithm}
\end{comment}

% \chen{May need a (sub)section to tell about the detailed implementation (e.g., firebase for UTG collection, framework or libary for image processing etc.)}
\section{Automated Evaluation of \tool}
\label{sec:evaluation}
In this section, we describe the procedure we used to evaluate \tool in terms of its performance automatically.
%The goal of our empirical study is to assess the accuracy, and performance of the approach.
We manually construct a dataset as the groundtruth for evaluating each step within our approach, instead of using the real-world bug recordings due to two reasons.
First, many real-world bug reports have been fixed and the app is also patched, but it is hard to find the corresponding previous version of the app for reproduction.
Second, the replay of some bug reports (e.g., financial, social apps) requires much information like authentication/database/hardware to generate the UTG which are beyond the scope of this study.
Therefore, we collect 61 visual recordings from 31 open-source Android apps which were used in previous studies~\cite{bernal2020translating,chaparro2019assessing,moran2015auto}.
They are also top-rated on Google Play covering 14 app categories (e.g., development, productivity, etc.).
To make these recordings as similar to real-world bug reports as possible, we adopt different ways for generating the recording including different creation tools (32 from video conversion, 22 from mobile apps, 7 from emulator screen recording), varied resolutions (27 $1920 \times 1080$, 23 $1280 \times 800$, 11 $900 \times 600$), diverse length (30-305 frames), and differed playing speed (7-30 frames per second).

For each app, we also collect its UTG by Google Firebase.
Since our approach consists of three main steps, we evaluate each phase of \tool, including Keyframe Location (Section~\ref{ref:phase1}), GUI Mapping (Section~\ref{ref:phase2}), and Execution Trace Generation (Section~\ref{sec:executiontrace}).
Therefore, we ask two experienced developers to manually label keyframes from recordings, GUI mapping between recording and UTG, and real trace in the UTG as the groundtruth for each phase. 
%\chen{Need to discuss the potential threat of data labeling.}
% \footnote{We will release the detailed dataset once paper acceptance.}
Note that each human annotator finished the labelling individually and they discussed the difference until an agreement was reached. 
%The detailed dataset can be seen

\begin{comment}
As the real-world visual bug recording may consist of GUIs that are not explored in our collected UTG (i.e., due to version updating), we construct an experimental dataset for automated evaluation, including 61 visual recordings from 31 open-source Android apps. %\chen{The size of data may be attacked by reviewers. Need another automated way in the future to easily collect large-scale dataset.} \sidong{I have added this threat in Section~\ref{sec:limitation}. Please confirm.}
We selected these apps because they have been used in previous studies~\cite{bernal2020translating,chaparro2019assessing,moran2015auto}, support different app categories and have a relatively high coverage UTG (i.e., 51.06 GUIs).
Since visual bug recordings are not readily available in these apps' issue trackers, we simulated the recordings according to our observation in Section~\ref{sec:observation} to ensure our simulated dataset reflects actual usage.
There are (i) 32:22:7 for creation methods (i.e., video convertion, mobile app, and emulator screen recording), (ii) 27:23:11 for different resolutions (i.e., $1920 \times 1080$, $1280 \times 800$, $900 \times 600$), and (iii) 12:41:8 for different GUI transition speed (i.e., quick, normal, and slow).
\end{comment}

%The main quality of our study is the extent to which GIFdroid can generate replayable execution trace that trigger the bug.

\subsection{Accuracy of Keyframe Location}
\label{sec:evaluation_1}
\textbf{Ground Truth:} 
% \chen{Need to give more details about the procedures in building the ground truth e.g., two participants do it individually and discuss the disagreement...}
% \sidong{Please confirm}
To evaluate the ability of keyframe location to accurately identify the keyframes present in the visual recordings, we manually generated a ground truth for the keyframe.
We recruited two paid annotators from online posting who have experience in bug replay.
To help ensure the validity and consistency of the ground truth, we first asked them to spend twenty minutes reading through the video annotation guidelines (i.e., TRECVID~\cite{over2013trecvid}) and get familiar with the widely-used annotation tool VirtualDub~\cite{web:virualdub}.
Then, we assigned the set of visual recordings to them to annotate keyframes independently without any discussion.
%Different from other tasks, the keyframe is not distinct due to the fact that an exactly same frame can last for several seconds.
%Therefore, we asked them to annotate a stable non-transient interval that involves a key activity, as the ground truth of a keyframe.
After the annotation, the annotators met and sanity corrected the subtle discrepancies (i.e., $\pm$ 3 frames).
Any disagreement would be handed over to one author for the final decision.
%Note that some keyframe annotations may encounter large discrepancies such as missing keyframe (one annotator had annotated the keyframe and one did not), large interval discrepancies ($>$ 3 frames).
%To decide such discrepancies rigorously, we asked one non-author Computer Science professor to make the final judgement.
In total, we obtained 289 keyframes for 61 recordings, 4.73 per recording on average, tallying with our observations for real-world recordings in Section~\ref{sec:general_statistics}.

\textbf{Metrics:} 
We employ three evaluation metrics, i.e., precision, recall, F1-score, to evaluate the performance of keyframe location.
The predicted frame which lies in the ground truth interval is regarded as correctly predicted.
Since there should be only one keyframe per interval, if two or more keyframes are localized in a single interval, only the first keyframe is counted as correct.
Precision is the proportion of frames that are correctly predicted as keyframes among all frames predicted as keyframes.
$$ precision = \frac{\#Frames \ correctly \ predicted \ as \ keyframes}{\#All \ frames \ predicted \ as \ keyframes} \\ $$
Recall is the proportion of frames that are correctly predicted as keyframes among all keyframes.
$$ recall = \frac{\#Frames \ correctly \ predicted \ as \ keyframes}{\#All \ keyframes} \\  $$
F1-score (F-score or F-measure) is the harmonic mean of precision and recall, which combine
both of the two metrics above.
$$ F1-score = \frac{2 \times precision \times recall}{precision + recall} $$
For all metrics, a higher value represents better performance.

\begin{comment}
\begin{equation}
\begin{gathered}
  precision = \frac{\#Frames \ correctly \ predicted \ as \ keyframes}{\#All \ frames \ predicted \ as \ keyframes} \\
  recall = \frac{\#Frames \ correctly \ predicted \ as \ keyframes}{\#All \ keyframes} \\ 
  F1-score = \frac{2 \times precision \times recall}{precision + recall}
  \label{eq:metric}
\end{gathered}
\end{equation}
\end{comment}

\textbf{Baselines:} We set up four state-of-the-art methods which are widely used for keyframe extraction as the baselines to compare with our method. \textit{Comixify}~\cite{pkesko2019comixify} is an unsupervised reinforcement learning method to predict for each frame a probability to extract keyframes in the videos.
\textit{ILS-SUMM}~\cite{shemer2019ils} formulates the keyframe extraction as an optimization problem, using a meta-heuristic optimization framework to extract a sequence of keyframes that has minimum feature distance.
\textit{Hecate}~\cite{song2016click} is a tool developed by Yahoo that estimates frame quality on the image aesthetics and clusters them to select the centroid as a keyframe.
\textit{PySceneDetect}~\cite{web:pyscenedetect} is a practical tool implemented as a Python library that is popular in GitHub.
The core technique for PySceneDetect to detect keyframes is a content-aware detection by analyzing the color, intensity, motion between frames.

\begin{comment}
\textit{Comixify}~\cite{pkesko2019comixify} is an unsupervised reinforcement learning method to extract keyframes in the videos.
It utilizes a deep summarization network (DSN) that predicts for each frame a probability, which indicates how likely a frame is to be selected as a representative frame of the video. 
%It outperforms on the task of keyframe extraction.

\textit{ILS-SUMM}~\cite{shemer2019ils} formulates the keyframe extraction as an optimization problem.
It first represents each frame as a feature and then uses the meta-heuristic optimization framework Iterated Local Search (ILS) to extract a sequence of keyframes that has minimum feature distance.

\textit{Hecate}~\cite{song2016click} is a tool developed by Yahoo to generate keyframe sequences from the video.
Hecate estimates frame quality using image aesthetics.
It clusters frames by their frame quality and selects a frame closest to the centroid as the most representative keyframe, one per cluster.

\textit{PySceneDetect}~\cite{web:pyscenedetect} is a practical tool implemented as a Python library that is popular in GitHub.
The core technique for PySceneDetect to detect keyframes is a content-aware detection by analyzing color, intensity, motion between frames.
\end{comment}

\begin{table}
\centering
\caption{Performance comparison for keyframe location.}
\label{tab:performance_rq1}
\begin{tabular}{|c|c|c|c|} 
\hline
\bf{Method} & \bf{Precision} & \bf{Recall} & \bf{F1-score}  \\ 
\hline
Hecate~\cite{song2016click} & 0.174 & 0.247 & 0.204  \\
Comixify~\cite{pkesko2019comixify} & 0.244 & 0.516 & 0.332 \\
ILS-SUMM~\cite{shemer2019ils} & 0.255 & 0.685 & 0.371 \\
PySceneDetect~\cite{web:pyscenedetect} & 0.418 & 0.550 & 0.475 \\
\hline
\bf{\tool} & \bf{0.858} & \bf{0.904} & \bf{0.880} \\
\hline
\end{tabular}
\end{table}

\textbf{Results:}
Table~\ref{tab:performance_rq1} shows the performance of all approaches.
The performance of our method is much better than that of other baselines, i.e., 32\%, 106\%, 14\% boost in recall, precision, and F1-score compared with the best baseline (ILS-SUMM, PySceneDetect).
The issues with these baselines are that they are designed for general videos which contain more natural scenes like human, plants, animals etc.
However, different from those videos, our visual bug recordings belong to artificial artifacts with different rendering processes. 
%\chen{Any more convincing points?}.
Therefore, considering the characteristics of visual bug recordings, our approach can work well in extracting keyframes. 
\begin{comment}
This implies that our method is especially good at keeping the keyframe from the visual recordings, i.e., larger improvement in recall.
Compared with the highest F1-score (PySceneDetect), our method gains about 41\% increase, indicating our method can locate a concise sequence of keyframes with higher accuracy.
This is due to these methods aim to extract keyframes for the natural scene.
In contrast, we aim to locate the frame that is fully rendered in the GUI visual artifacts, a tailored keyframe location method is more suitable.
\end{comment}

Our approach also makes mistakes in keyframe extraction due to three reasons.
First, within some apps, the resource loading is so slow that the partial GUI may stay for a relatively long time, beyond our threshold setting in Section~\ref{ref:phase1}.
So, that frame is wrongly predicted as a keyframe.
Second, some users may click the button before the GUI is fully rendered to the next page.
That short period makes our approach miss the keyframe.
Third, some GUIs may contain animated app elements such as advertisement or movie playing, which will change dynamically, resulting in no steady keyframes being localized.

\begin{comment}
\sidong{Instances that we made mistakes are mainly due to two reasons.
First, the resources in the GUIs are too slow to render, as it takes longer than our empirical threshold in Section~\ref{ref:phase1}, which leads our method to locate an incompletely rendered GUI as a keyframe.
Second, GUI transitions are too fast or GUIs contain dynamic elements, causing our method can not capture a steady state.}
\end{comment}

\subsection{Accuracy of GUI Mapping}
\label{sec:exp_phase2}
\textbf{Ground Truth:}
To evaluate the ability of GUI mapping to accurately search the nearest GUI screenshot in the UTG, we first collected the UTG and their GUI screenshots for those apps, following the procedure outlined in Section~\ref{sec:utg_collection}.
Given the keyframes (i.e., a frame from the ground truth interval), we selected the most similar GUI screenshot in the UTG as the ground truth for GUI mapping.
Note that labelling was finished by two annotators independently without any discussion, and any disagreement would be handed over to one author for the final decision.
Thus, we obtained 289 ground truth pairs of keyframes and GUI screenshots.

\begin{table*}
	\centering
	\caption{Performance comparison for GUI Mapping}
	\label{tab:performance_rq2}
	\begin{tabular}{|c|c|c|c|c|c|c|c|c|c|c|} 
		\hline
		\multirow{2}{*}{\bf{Metric}} & \multicolumn{5}{c|}{Pixel} & \multicolumn{4}{c|}{Stucture} & \tool \\ 
		\cline{2-11}
		& Euclidean & Histogram & pHash & dHash & aHash & SSIM & SIFT & SURF & ORB & SSIM+ORB \\ 
		\hline
		\bf{Precision@1} & 0.0\% & 30.6\% & 46.6\% & 50.6\% & 38.6\% & 75.7\% & 63.6\% & 67.8\% & 75.7\% & \bf{85.4\%} \\
		\bf{Precision@2} & 1.3\% & 47.7\% & 58.6\% & 67.8\% & 61.3\% & 81.3\% & 81.3\% & 77.3\% & 83.7\% & \bf{90.0\%} \\
		\bf{Precision@3} & 1.3\% & 55.7\% & 63.6\% & 75.7\% & 66.6\% & 85.3\% & 89.3\% & 82.6\% & 89.3\% & \bf{91.3\%}\\
		\hline
	\end{tabular}
	\vspace{0.3cm}
\end{table*}

\textbf{Metrics:}
We formulate the problem of GUI mapping as an image searching task (i.e., search the most similar GUI screenshot), so we adopt Precision@k to evaluate the performance of GUI mapping.
The higher value of the metric is, the better a search method performs. 
Precision@k is the proportion of the top-k results for a query GUI that contains the grouptruth one.
Note we only consider $k$ in the range 1-3, as developers rarely check a long recommendation list.

\begin{comment}
\begin{equation}
	Precision = \frac{\text{No. images mapped}} 
	{\text{Total No. images mapped}}
\end{equation}
Since Top-1 precision may not map all desired information~\cite{salton1992state}, we evaluate the precision at particular ranks (Precision@1, Precision@2, Precision@3).
Precision@K means the precision after the top-K mapped images.
The higher the metric score, the more images are mapped to the ground truth.

\end{comment}
\textbf{Baselines:}
To further demonstrate the advantage of our method, we compare it with 10 image processing baselines, including pixel level (e.g, euclidean distance~\cite{danielsson1980euclidean}, color histogram~\cite{wang2010robust}, fingerprint~\cite{alsmirat2019impact}), and structural level (e.g., SSIM~\cite{wang2004image}, SIFT~\cite{lowe2004distinctive}, SURF~\cite{bay2006surf}, ORB~\cite{rublee2011orb}).

\begin{figure}  
	\centering 
	\includegraphics[width=0.95\linewidth]{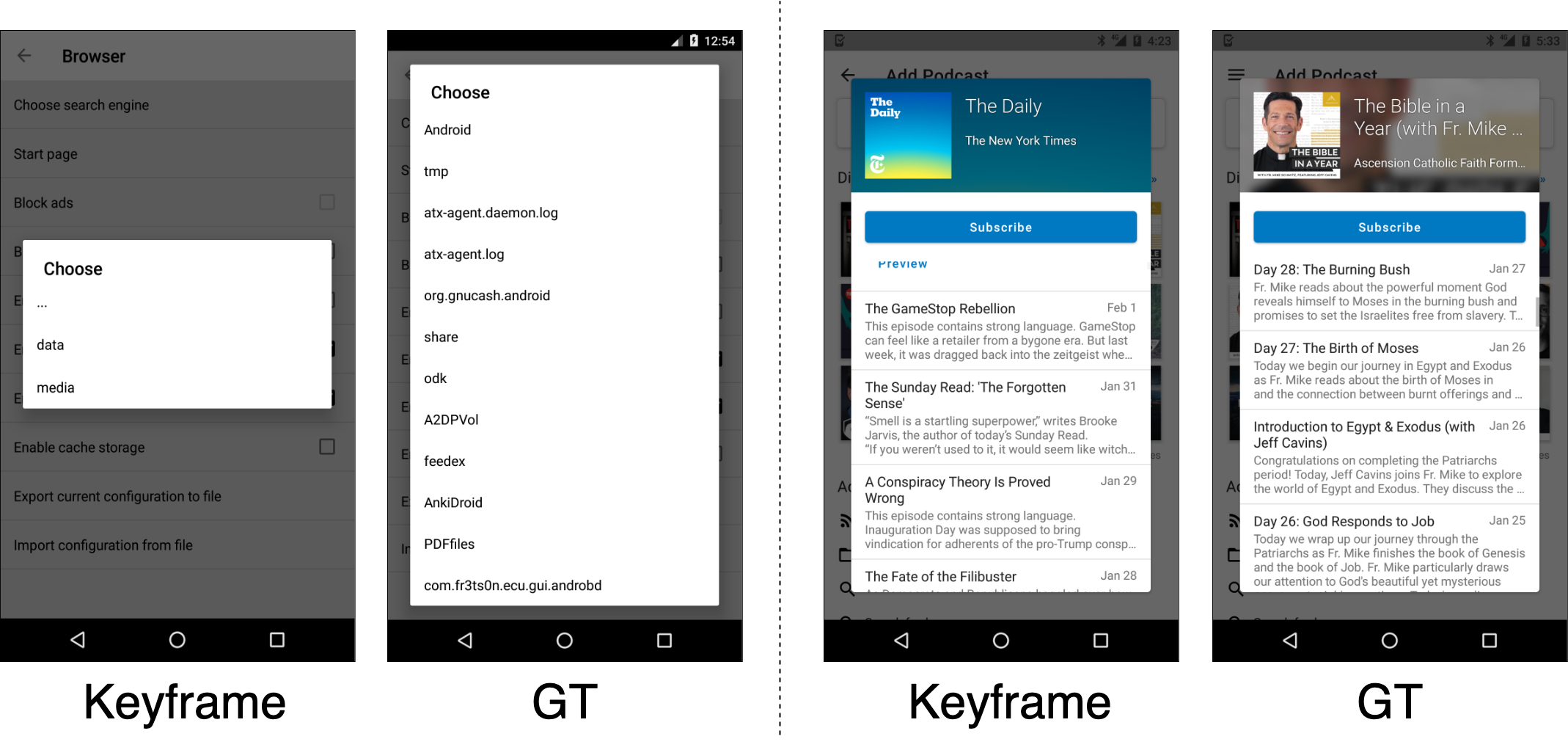} 
	\caption{Examples of bad cases in GUI mapping.} 
	\label{fig:wrong_mapping}
\end{figure}

\textbf{Results:}
Table~\ref{tab:performance_rq2} shows the overall performance of all methods.
In contrast with baselines, our method outperforms in all metrics, 85.4\%, 90.0\%, 91.3\% for Precision@1, Precision@2, Precision@3 respectively.
We observe that the methods based on structural features perform much better than pixel features due to the reason that the pixel similarity suffers from the scale-invariant as the resolutions for visual recordings varies.
Our method that combines SSIM and ORB leads to a substantial improvement (i.e., 9.7\% higher) over any single feature, indicating that they complement each other.
In detail, ORB addresses the image distortion that causes false GUI mapping considering only SSIM.
%On the one hand, SSIM addresses the limitations of few features detected in ORB method.
%On the other hand, ORB addresses the image distortion that causes falsely GUI mapping in SSIM method.

Albeit the good performance of our method, we still make wrong mapping for some keyframes.
We manually check those wrong mapping cases, and find that one common cause is the dynamically loaded content.
For example, as seen in Figure~\ref{fig:wrong_mapping}, some GUIs mappings look visually very different as the pop-up window or images loaded from the internet may vary a lot at different time.
As the dynamic content takes over a large area of the whole GUI, our approach based on visual features cannot accurately locate them.
%that a layout rendering different content. The wrongly mapping cases particularly occur in the UI consists of large area of content as shown in Figure~\ref{fig:wrong_mapping}. We discuss potential solutions to this limitation in Section~\ref{sec:limitation}.
\subsection{Performance of Trace Generation}
\label{sec:eval_rq3}
\textbf{Ground Truth:}
To evaluate the ability of trace generation to accurately generate the replayable execution trace, we label the execution trace that can replay the visual recording from the app launch as the ground truth.
Note that there may be multiple ground truth traces complying with the optimal definition (i.e., the shortest trace to replay the recording) in Section~\ref{sec:executiontrace}.
Therefore, we manually check and label all the ground truths by two annotators independently without any discussion.
To ensure the quality, any disagreement would be handed over to one author for the final decision.
Totally, we obtained 61 execution traces, including 539 reproduction steps.

\textbf{Metrics:}
To measure the similarity of the groundtruth trace and predicted replay sequence, we first get the LCS between two sequences.
We then calculate the similarity~\cite{wolk2014sentence} following $\frac{2 \times M}{T}$ where $M$ is the length of LCS, and $T$ is the length of the sum of both sequences.
%adopt the \url{https://docs.python.org/3/library/difflib.html} to calculate the similarity between two sequences.
%\sidong{It is defined as the measurement between LCS and their sequence length: $2 \times M / T$ where $M$ is the number of LCS, and $T$ is the total number of elements in both sequences.}
It gives a real value with a range [0,1] and is usually expressed as a percentage. 
The higher the metric score, the more similar the generated execution trace is to the ground truth.
If the generated trace exactly matches the ground truth, the similarity value is 1 (100\%).

\textbf{Baselines:}
We set up one state-of-the-art scenario generation method (V2S) and an ablation study of GIFdroid without LCS algorithm (GIFdroid-LCS) as our baselines.
\textit{V2S}~\cite{bernal2020translating} is the first purely graphical Android record-and-replay technique.
V2S adopts deep learning models to detect user actions via classifying touch indicators for each frame in a video, and converts these into a script to replay the scenarios.
Note that V2S has strict requirements about the input video including high resolution, frames per second and touch indicator for inferring detailed actions.
Therefore, in addition to testing V2S in all our datasets, we also test its performance in the part of our dataset (i.e., 19 recordings) which contain touch indicators  called \textit{V2S+Touch}.
%Since our dataset contains the recordings with mobile touch indicator, mouse, and non touch indicator, we also set up a baseline \textit{V2S+Touch} to evaluate V2S on mobile touch indicator only dataset \sidong{(i.e., 19 recordings)}.
% Due to its specification of requirement on touch indicator, we also set up two derivations for evaluation.
% \textit{V2S} evaluates the V2S for the general dataset which contains mobile touch indicator, mouse, or non touch indicator.
% \textit{V2S+Touch} evaluates the V2S for the recordings that contains mobile touch indicator only.
To test the importance of the proposed algorithm that addresses defective index sequence in Section~\ref{sec:executiontrace}, we also add another ablation study called \textit{GIFdroid-LCS} which does not contain that component.
%It generates the execution trace by combining the anterior trace (i.e., using Depth-First Search to search a trace from app launch to the first index), and the index trace (i.e., using keyframe location and GUI mapping).

\begin{table}
	\centering
	\caption{Performance comparison of execution generation.}
	\label{tab:performance_rq3}
	\begin{tabular}{|l|c|c|c|c|} 
		\hline
		\bf{Metric} & V2S & V2S+Touch & GIFdroid-LCS & \tool \\ 
		\hline
		\bf{Similarity} & 7.17\% & 63.19\% & 82.63\% & \bf{89.59\%} \\
		\bf{Time (sec)} & 791.23 & 856.91 & \bf{84.66} & 97.91 \\
		\hline
	\end{tabular}
	\vspace{0.2cm}
\end{table}

\begin{table*}
	\centering
% 	\footnotesize
	\caption{Detailed results for execution trace generations. Green cells indicate fully reproduced recordings, orange cells $>50\%$ reproduced, and red cells $<50\%$.}
	\label{tab:performance_rq3_detail}
	\begin{tabular}{|l|c|c|l|c|c|l|c|c|l|c|c|} 
		\hline
		\rowcolor{Gray}
		\bf{AppName} & \multicolumn{2}{c|}{\bf{Rep. Trace}} & \bf{AppName} & \multicolumn{2}{c|}{\bf{Rep. Trace}} & \bf{AppName} & \multicolumn{2}{c|}{\bf{Rep. Trace}} & \bf{AppName} & \multicolumn{2}{c|}{\bf{Rep. Trace}} \\ 
		\hline
		Token & \cellcolor{green} 8/8 & \cellcolor{green} 5/5 &
		TimeTracker & \cellcolor{green} 10/10 & \cellcolor{green} 8/8 & 
		TrackerControl & \cellcolor{orange} 23/24 & \cellcolor{green} 12/12 &
		YoloSec & \cellcolor{green} 9/9 & \cellcolor{orange} 8/10
		\\
		\hline
		DeadHash & \cellcolor{green} 25/25 & \cellcolor{green} 18/18 &
		GNUCash & \cellcolor{green} 9/9 & \cellcolor{green} 7/7 &
		aFreeRDP & \cellcolor{green} 11/11 & \cellcolor{green} 10/10 & 
		AntennaPod & \cellcolor{green} 7/7 & \cellcolor{orange} 8/9
		\\
		\hline
		ProtonVPN & \cellcolor{green} 7/7 & \cellcolor{green} 9/9  &
		FastNFitness & \cellcolor{green} 8/8 & \cellcolor{orange} 8/10 & 
		JioFi & \cellcolor{green} 6/6 & \cellcolor{green} 8/8 &
		WiFiAnalyzer & \cellcolor{green} 9/9 & \cellcolor{green} 5/5
		\\
		\hline
		PSLab & \cellcolor{orange} 8/9 & \cellcolor{green} 6/6 &
		DroidWeight & \cellcolor{green} 6/6 & \cellcolor{green} 7/7 & 
		openScale & \cellcolor{green} 18/18 & \cellcolor{ored} 3/18 &
		KeePassDX & \cellcolor{green} 6/6 & \cellcolor{green} 7/7
		\\
		\hline
		Trigger & \cellcolor{green} 7/7 & \cellcolor{green} 12/12 &
		ADBungFu & \cellcolor{ored} 2/8 & \cellcolor{green} 21/21 & 
		EteSyncNotes & \cellcolor{green} 11/11 & \cellcolor{green} 12/12 &
		PortAuthority & \cellcolor{green} 8/8 & \cellcolor{green} 12/12
		\\
		\hline
		ATime & \cellcolor{green} 7/7 & \cellcolor{green} 8/8 &
		AdAway & \cellcolor{green} 11/11 & \cellcolor{green} 19/19 & 
		StinglePhoto & \cellcolor{green} 7/7 & \cellcolor{green} 6/6 &
		InviZible & \cellcolor{green} 6/6 & \cellcolor{green} 10/10
		\\
		\hline
		% F-DroidBuild & \cellcolor{orange} 7/11 & \cellcolor{green} 6/6
	\end{tabular}
% 	\vspace{0.5cm}
\end{table*}

\begin{figure}
	\centering
	\includegraphics[width=0.95\linewidth]{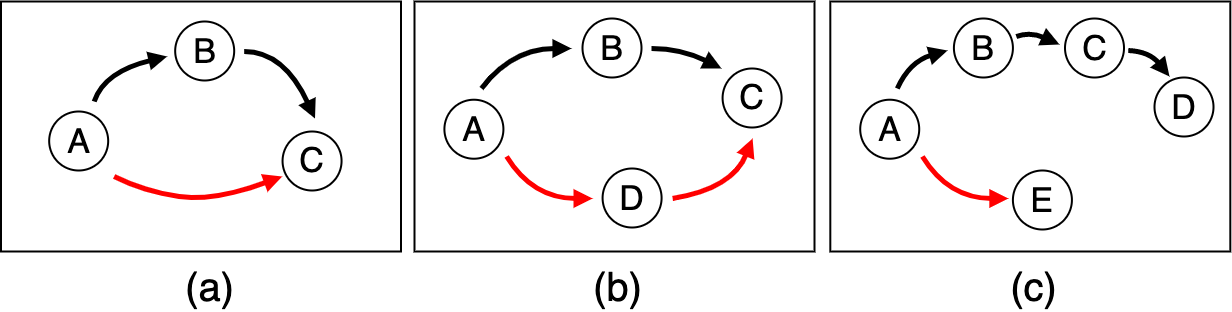}
	\caption{Illustration of bad case for trace generation. Black line represents ground truth. Red line represents the generated execution trace.}
	\label{fig:badcase_rq3}
% 	\vspace{0.2cm}
\end{figure}

\textbf{Results:}
Table~\ref{tab:performance_rq3} shows the performance comparison with the baselines.
Our method achieves 89.59\% sequence similarity which is much higher than that of baselines.
Note that due to the strict requirement of input recordings, V2S does not work well in all our datasets, but performs well in our partial dataset with touch indicators.
Even for recordings with touch indicators, the extracted trace is still not that accurate as it could not recover the trace from the app launch to the entry in the recording.
In a word, the hard requirements still limit its generality in the real testing environment especially those open-source software development.
\begin{comment}
A large gap between V2S and V2S+Touch indicates that V2S is not compatible in general recordings (i.e., usually does not have touch indicators). 
In addition, V2S+Touch achieves a relatively low similarity value (63.19\%) due to its limitation on generating anterior trace from the app launch to the first keyframe.
\end{comment} 
In addition, adding LCS can mitigate the errors introduced in the first two steps in our approach, resulting in a boost of performance from 82.63\% to 89.59\%.
% is due to the usage of proposed algorithm (LCS), revealing the ability of \tool to solve the challenges of defective index sequence.
Although applying LCS takes a bit more runtime (i.e., 13.25 seconds on average), it does not influence its real-world usage as it can be automatically run offline.
% this process is fully automated, can run in the background, and can be accelerated by more advanced algorithm.

Table~\ref{tab:performance_rq3_detail} shows detailed results of the success rate for each visual recordings, where each app, execution trace (number of steps), and successfully replayed steps are displayed.
Green cells indicate a fully reproduced execution trace from the video recording, Orange cells indicate more than 50\% of steps reproduced, and Red cells indicate less than 50\% of reproduced steps. 
\tool fully reproduces 82\% (50/61) of the visual recordings, signals a strong replay-ability.
We manually check the instances where our method failed to reproduce scenarios.
Instances where slightly biased (orange cells) are largely due to inaccuracies in keyframe location (i.e., in Figure~\ref{fig:badcase_rq3}(a), $\{B\}$ is missing) and GUI mapping (i.e., in Figure~\ref{fig:badcase_rq3}(b), $\{B\}$ is incorrectly mapping to $\{D\}$).
Instances that failed to reproduce (red ceils) are due to the inaccuracies on the last index as our method depends on the last index to end the search (i.e., in Figure~\ref{fig:badcase_rq3}(c), we search the execution trace on $\{E\}$).

% First, our method may miss a keyframe in a connected node, for example as shown in Fig.~\ref{fig:badcase_rq3}(a), due to the keyframe $\{B\}$ is unidentified, the output of our approach ($\{A,C\}$) is slightly bias to the ground truth.
% Second, our method may incorrectly map to nodes that also have associations.
% For example, in Fig.~\ref{fig:badcase_rq3}(b), due to node $\{B\}$ is incorrectly mapping to node $\{D\}$, the ground truth $\{A,B,C\}$ is bias to $\{A,D,C\}$.
% Instances where failed to reproduce (red ceils) are due to the inaccuracies on last index in GUI mapping as our method depends on the last index to end the search.
% For example, in Fig.~\ref{fig:badcase_rq3}(c), once the last keyframe is incorrectly mapping to node $\{E\}$, our approach fails to reproduce the ground truth trace.
%We discuss potential solutions to this limitation in Section~\ref{sec:limitation}.

\section{Usefulness Evaluation}
We conduct a user study to evaluate the usefulness of the generated execution trace for replaying visual bug recording into real-world development environments.
We recruit 8 participants including 6 graduate students (4 Master, 2 Ph.D) and 2 software developers to participate in the experiment.
All students have at least one-year experience in developing Android apps and have worked on at least one Android apps project as interns in the company.
Two software developers are more professional and work in a large company (Alibaba) about Android development.

\textbf{Procedure:}
\begin{comment}
\sidong{We start with a basic demographic and technical survey to understand the participantsâ background. 
All of the graduate students (4 Master, 2 Ph.D) are studying in the major of software engineering at top universities in different countries.
They all have at least one year experience in developing Android apps and have worked on at least one Android apps project as intern in the company, suggesting them as sufficient developer proxies in controlled experiments~\cite{salman2015students}.
And two software developers are working in large companies with two year working experiences in Android development, while one focuses on developing software solutions to meet user needs and the other is responsible for testing and debugging app infrastructure.}
\end{comment}
% All of the graduate students are studying in the major of software engineering at top universities, and all of the software developers are working in large companies with two year experience in Android apps development.
% According to the pre-study background survey, all of the recruited participants have experience in developing Android apps for more than one year. 
We first give them an introduction to our study and also a real example to try.
Each participant is then asked to reproduce the same set of 10 randomly selected visual bug recordings from GitHub which are of diverse difficulty ranging from 6 to 11 steps until triggering bugs.
%\sidong{real world visual bug reports from GitHub} 
% including three relatively easy to reproduce (1-7 steps), four medium (7-9 steps), and two hard ($>$9 steps) for the study.
%To further demonstrate the generality of our recordings, we show the examples of real-world visual recordings and the generated execution traces by GIFdroid in our online appendix\footnote{https://sites.google.com/view/gifdroid}.
Detailed experiment dataset can be seen in our online appendix\footnote{\url{https://sites.google.com/view/gifdroid}}.
The study involves two groups of four participants: the experimental group $P_1$, $P_2$, $P_3$, $P_4$ who gets help with the generated execution trace by our tool, and the control group $P_5$, $P_6$, $P_7$, $P_8$ who starts from scratch.
Each pair of participants $\langle P_x$ , $P_{x+4}\rangle$ have comparable development experience, so that the experimental group has similar capability to the control group in total.
Note that we do not ask participants to finish half of the tasks with our tool while the other
half without assisting tool to avoid potential tool bias.
Our tool takes about 138.08s on average to generate execution trace for the average 10.59s bug recording as seen in the Table~\ref{tab:efficiency}.
We only record the time used to reproduce the visual bug recordings in Android, as the execution trace generation can be finished offline, especially for long recordings which require much processing time. Since our approach is fully automated, our model can automatically deal with the bug video recording immediately once uploaded.
Participants have up to 10 minutes for each bug replay.

\begin{table*}
\centering
\small
\caption{The execution time of \tool for real-world bug recordings}
\label{tab:efficiency}
\begin{tabular}{|l|c|c|c|c|c|} 
\hline
\multirow{2}{*}{\bf{AppName}} & \bf{Gif} & \multicolumn{4}{c|}{\bf{\tool Execution}} \\ 
\cline{2-6}
 & \bf{Duration (sec)} & \bf{Keyframe Location (sec)} & \bf{GUI Mapping (sec)} & \bf{Trace Generation (sec)} & \bf{Total (sec)} \\ 
\hline
    AnkiDroid-1 & 16.7 & 28.05 & 174.96 & 0.58 & 203.59 \\
\hline
    AnkiDroid-2 & 9.5 & 13.95 & 84.80 & 1.12 & 99.87 \\
\hline
    KISS-1 & 13.8 & 16.74 & 148.51 & 0.13 & 165.38 \\
\hline
    NeuroLab-1 & 9.6 & 14.76 & 71.05 & 0.01 & 85.82 \\
\hline
    NeuroLab-2 & 4.5 & 4.76 & 77.32 & 0.13 & 82.21 \\
\hline
    BeHe-1 & 11.8 & 7.10 & 168.77 & 0.11 & 175.98 \\
\hline
    BeHe-2 & 6.0 & 3.73 & 64.80 & 0.60 & 69.13 \\
\hline
    AndrOBD-1 & 7.9 & 10.00 & 115.56 & 0.41 & 125.97 \\
\hline
    PDFConverter-1 & 17.4 & 36.22 & 191.03 & 2.32 & 229.57 \\
\hline
    PDFConverter-2 & 8.7 & 9.99 & 131.51 & 1.84 & 143.34 \\
\hline
\hline
Average & 10.59 & 14.53 & 122.83 & 0.72 & 138.08 \\
\hline
\end{tabular}
\end{table*}

\begin{table}
\centering
\small
\caption{Performance comparison between the experimental and control group. $^*$ denotes \textit{p} $<$ 0.01.}
\label{tab:performance_rq4}
\begin{tabular}{|l|c|c|c|c|} 
\hline
\multirow{2}{*}{\bf{AppName}} & \multicolumn{2}{c|}{\bf{Control Group}} & \multicolumn{2}{c|}{\bf{Experimenal Group}} \\ 
% \cmidrule(){2-5}
\cline{2-5}
 & \bf{Success} & \bf{Time (sec)} & \bf{Success} & \bf{Time (sec)} \\ 
\hline
    AnkiDroid-1 & 2/4 & 473 & 4/4 & 102 \\
\hline
    AnkiDroid-2 & 3/4 & 334 & 4/4 & 95 \\
\hline
    KISS-1 & 4/4 & 121 & 4/4 & 82 \\
\hline
    NeuroLab-1 & 4/4 & 90 & 4/4 & 73 \\
\hline
    NeuroLab-2 & 4/4 & 366 & 4/4 & 62 \\
\hline
    BeHe-1 & 4/4 & 77 & 4/4 & 55 \\
\hline
    BeHe-2 & 4/4 & 94 & 4/4 & 51 \\
\hline
    AndrOBD-1 & 4/4 & 63 & 4/4 & 56 \\
\hline
    PDFConverter-1 & 4/4 & 71 & 4/4 & 47 \\
\hline
    PDFConverter-2 & 4/4 & 25 & 4/4 & 24 \\
\hline
\hline
Average & 3.7/4 & 171.4 & 4/4 & 65.0$^*$ \\
\hline
\end{tabular}
\vspace{0.2cm}
\end{table}

\textbf{Results:}
Table~\ref{tab:performance_rq4} shows the experiment result.
Although most participants from both experimental and control groups can successfully finish the bug replay on time, the experiment group reproduces the visual bug recording much faster than that of the control group (with an average of 171.4 seconds versus 65.0 seconds). 
In fact, the average time of the control group is underestimated, because three bugs fail to be reproduced within 10 minutes, which means that participants may need more time. 
In contrast, all participants in the experiment group finish all the tasks within 2 minutes.
To understand the significance of the differences between the two groups, we carry out the Mann-Whitney U test~\cite{mann1947test} (specifically designed for small samples) on the replaying time.
The testing result suggests that our tool can significantly help the experimental group reproduce bug recordings more efficiently ($p−value<0.01$).

We summarise two reasons why it takes the control group more time to finish the reproduction than the experiment group.
First, some visual recording is quite complicated which requires participants in the control group to watch the visual recordings several times for following procedures.
The GUI transitions within the recording may also be too fast to follow, so developers have to replay it.
Second, it is hard to determine the trigger from one GUI to the next one.
As illustrated in Section~\ref{sec:general_statistics}, only 25\% of videos are recorded with the touch indicator, resulting in developers' guess of the action for triggering the next GUI.
That trial and error makes the bug replay process tedious and time-consuming.
It is especially severe for junior developers who are not familiar with the app code.
% \chen{Any qualitative feedback from developers about our tool?}
%some participants in the control group need multiple attempts to find the entry between GUIs.
%It appears not only when finding the entry of the visual recording, but also when reproducing the steps in the recording as some recordings do not indicate the user behaviour.

\vspace{0.2cm}

\begin{comment}
\begin{table}
\centering
\footnotesize
\caption{Detailed Results for RQ4 real world evaluation.}
\label{tab:performance_rq4}
\begin{tabular}{|l|c|c|l|c|c|} 
\hline
\rowcolor{Gray}
\bf{AppName} & \multicolumn{2}{c|}{\bf{Rep. Trace}} & \bf{AppName} & \multicolumn{2}{c|}{\bf{Rep. Trace}} \\ 
\hline
    AnkiDroid & \cellcolor{green} 9/9 & \cellcolor{green} 11/11 &
    PDFConverter & \cellcolor{green} 11/11 & \cellcolor{green} 7/7
    \\
\hline
    BeHe & \cellcolor{orange} 6/8 & \cellcolor{green} 6/6 &
    NeuroLab & \cellcolor{green} 8/8 & \cellcolor{green} 11/11
    \\
\hline
    AndrORB & \cellcolor{green} 8/8 & \cellcolor{lightgray} N/A &
    KISS & \cellcolor{green} 7/7 & \cellcolor{lightgray} N/A
    \\
\hline
\end{tabular}
\end{table}
\end{comment}
% \input{experiments}
% \input{results}

\section{Threats to validity}
We had discussed the limitations of our approach at the end of each subsection of the evaluation in Section~\ref{sec:evaluation}, such as errors due to slow rendering in keyframe location (Section~\ref{sec:evaluation_1}), dynamic loading content in GUI mapping (Section~\ref{sec:exp_phase2}), etc.
% Our approach has various limitations that serve as motivation for future work.
In this section, we further discuss the threats to validity of our approach.

\textbf{Internal Validity.}
In our automated experiments evaluating \tool, threats to internal validity may arise from human annotators and artificial recordings.
To help mitigate this threat, we recruited two paid annotators from online posting who have experience in video annotation. 
To mitigate any potential subjectivity or errors, we asked them to annotate independently without any discussion, and then met and sanity corrected the discrepancies.
% Furthermore, we have released all of our experimental data and code [26], to facilitate the reproducibility of our experiments.
Another potential threat concerns the selection of optimal execution trace. 
To mitigate this threat, we chose the shortest candidate sequence as the optimal execution trace to help developers reproduce the bug with the least amount of time/steps.

\textbf{External Validity.}
The main threat to external validity arises from the potential bias in the selection of experimental apps used in our automated evaluation.
To help mitigate this threat, we utilized the apps used in previous studies~\cite{bernal2020translating,chaparro2019assessing,moran2015auto}, which have undergone several filtering and quality control mechanisms to ensure diversity.
One more potential external threat concerns the generalizability on the manual creation of visual recordings for validating the performance of our approach. 
For example, there are many different types of devices with different screens especially in Android which may result in different GUI rendering.
To mitigate this threat, we took care to
generate the recordings as general as possible, including different creation tools, varied resolutions, diverse length and differed recording speed.

\section{Related Work}
% \chen{Since there are some new papers especially in FSE'21, ASE'21, Can you please have a rough look and add some of them accordingly?}
A growing body of tools has been dedicated to assisting in recording and replaying bugs in mobile and web apps.
We introduce related works of bug replay based on different types of information including the app running information, textual description, and visual recording in this section.

\subsection{Bug Record and Replay from App Running Information}
Nurmuradov et al.~\cite{nurmuradov2017caret} introduced a record and replay tool for Android applications that captures user interactions by displaying the device screen in a web browser. 
It used event data captured during the recording process to generate a heatmap that facilitates developers understanding of how users are interacting with an application. 
%This approach is limited in that users must interact with a virtual Android device through a web application, which could result in unnatural usage patterns.
Additional tools including ECHO~\cite{sui2019event}, Reran~\cite{gomez2013reran}, Barista~\cite{ko2006barista}, and Android Bot Maker~\cite{web:botmaker} are program-analysis related applications for the developer to record and replay interactions. 
However, they required the installation of underlying frameworks such as replaykit~\cite{web:replaykit}, troyd~\cite{jeon2012troyd}, or instrumenting apps which is too heavy for end users.
In contrast to these approaches, our \tool is rather light-weight which just requires the input recording and UTG which can be generated by existing tools.
% there are no frameworks to install or instrumentation to add, in our tool GIFdroid, making its ease of use for developers.

\subsection{Bug Replay from Textual Information}
To lower the usage barrier of record and replay that requires frameworks, many of the studies ~\cite{moran2015auto,bell2013chronicler,kifetew2014reproducing,fazzini2018automatically,white2015generating} facilitate the bug replay base on the natural language descriptions in bug report which contains immediately actionable knowledge (e.g., reliable reproduction steps), stack traces, test cases, etc.
% created by bug reporters (e.g., testers, beta users), 
For example, ReCDroid~\cite{zhao2019recdroid} leveraged the lexical knowledge in the bug reports and proposed a natural language processing (NLP) method to automatically reproduce crashes.
However, it highly depended on the description writing in the bug report including formatting, word usage, granularity~\cite{erfani2014works,bettenburg2008makes,koru2004defect} which limit its deployment in the real world. 
Many works that assist developers in writing bug reports~\cite{moran2015auto,mani2012ausum,moreno2017automatic,lotufo2015modelling} still cannot directly benefit it.
%However, the problem facing many current bug reporting systems is that typical natural language reports capture a coarse grained level of detail that makes developer reasoning about defects difficult~\cite{erfani2014works,bettenburg2008makes,koru2004defect}.
%Although there are numerous works focusing on improving and completing the bug reports~\cite{moran2015auto,mani2012ausum,moreno2017automatic,lotufo2015modelling}, visual recordings are more useful and easier to understand as bridging the lexical gap exists between reporters and developers~\cite{bernal2020translating}.
Different from textual bug reports, visual recordings contain more bug details and can be easily created by non-technical users.
Therefore, our work focuses on the automated bug replay from these visual bug reports.

\subsection{Bug Replay from Visual Information}
In software engineering, there are many works on visual artifacts including recording, tutorials, bug reports~\cite{ponzanelli2016too,frisson2016inspectorwidget,chen2018ui,chen2020wireframe,cooper2021takes,yu2021layout,zhao2021guigan} with some of them specifically for usability and accessibility testing~\cite{liu2020owl, chen2020unblind, zhao2020seenomaly, yang2021don}.
% have focused on generating the replay scenarios by using high-level descriptions of events from reporter video recording.
In detail, Bao et al.~\cite{bao2017extracting} focused on the extraction of user interactions to facilitate behavioral analysis of developers during programming tasks, such as code editing, text selection, and window scrolling.
%In our work, rather than focusing on understanding and extracting fine-grained developers interactions, we instead focus on helping with bug reproduction and generating a high fidelity replay execution traces on generic user actions on mobile apps.
Krieter et al.~\cite{krieter2018analyzing} analyzed every single frame of a video to generate log files that describe what events are happening at the app level (e.g., the "k" key is pressed).
Different from these works in extracting developers' behavior, our work is concerned with the automated bug replay from the recording in the bug report.
%However, our work focuses on video analysis to help with generation of replayable execution traces, rather than describing usage scenarios at a high level.

Bernal et al~\cite{bernal2020translating} proposed a tool named V2S which leverages deep learning techniques to detect and classify user actions (e.g., a touch indicator that gives visual feedback when a user presses on the device screen) captured in a video recording, and translate it into a replayable test script.
Although the target of this work is the same as ours, there are two major differences between them.
First, V2S requires high-resolution recording with touch indicators which is hard to get in real-world bug reports according to our analysis (less than 25.8\%) in Section~\ref{sec:observation}. 
Second, it requires complete recording from the app launch to the bug occurrence while most recordings (93.2\%) in the real world do not start from the app launch, but just 2-7 steps before the bug page. 
%Compared to our work, this technique is limited to produce a replay script for generic reporter video recording as only 25.8\% of video recording contained a touch indicator, and even fewer 6.8\% recorded from app launch as discussed in Section~\ref{sec:observation}.
In contrast, our approach design is fully driven by the practical data, hence it does not have those requirements.
The user study of real-world bug replay also confirms the usefulness and generality of our approach.
%uses light-weight image processing techniques to locate a sequence of keyframes that maps to the UTG of the app, and then adopt a novel algorithm to generate the execution replay trace, aiming to reproduce the bug for general visual recordings.

%\chen{IDEA: We may have one work in the future that can both record the screen and record \& replay.}

\vspace{0.3cm}
\section{Conclusion}
% \chen{May add some space between the figure/table and text to make the paper exactly of 10 full pages.}
The visual bug recording is trending in bug reports due to its easy creation and rich information.
To help developers automatically reproduce those bugs, we propose \tool, an image-processing approach to covert the recording to executable trace to trigger the bug in the Android app.
Our automated evaluation shows that our \tool can successfully reproduce 82\% (50/61) visual recordings from 31 Android apps.
The user study on replaying 10 real-world visual bug recordings confirms the usefulness of our \tool in boosting developers' productivity.

In the future, we will keep improving our method for better performance in term of keyframe extraction, GUI mapping, and trace generation.
For example, the traditional image processing methods may not robust to minor GUI changes such as configuration and parameter changes.
We will further improve our approach to locate more fine-grained parameter information.
To make \tool more usable, we will also take the human factor into the consideration.
As the automated approach may not be perfect, we will further explore how human can collaborate with the machine for replaying the bugs in the visual bug report.
While \tool is fully automated, can run in the background, the execution overhead is not ideal.
In the future, we will improve the efficiency of our approach, for example, accelerating by more advanced hardware and algorithm.

\bibliographystyle{ACM-Reference-Format}
\bibliography{main.bib}

\end{document}